\documentclass[pre,twocolumn,superscriptaddress,floatfix,amssymb,showpacs,aps,10pt]{revtex4-1}
\usepackage{enumerate}
\usepackage{times}
\usepackage{graphicx}
\usepackage{amsmath}
\usepackage{color}
\usepackage{natbib}
\usepackage{array}
\usepackage{float}
\newcolumntype{C}[1]{>{\centering\let\newline\\\arraybackslash\hspace{0pt}}m{#1}}
\usepackage[pdfpagemode=None,colorlinks=true,urlcolor=black,%
linkcolor=blue,citecolor=blue,pdfstartview=FitH]{hyperref}
\bibpunct{[}{]}{,}{n}{}{;}
\usepackage{lipsum}
\usepackage{ulem}
            
\graphicspath{{fig/}} 
\newcolumntype{L}[1]{>{\raggedright\let\newline\\\arraybackslash\hspace{0pt}}m{#1}}
\newcolumntype{C}[1]{>{\centering\let\newline\\\arraybackslash\hspace{0pt}}m{#1}}
\newcolumntype{R}[1]{>{\raggedleft\let\newline\\\arraybackslash\hspace{0pt}}m{#1}}

\begin{document}

\title{Dynamics of driven translocation of semiflexible polymers}

\author{P. M. Suhonen}
\author{R. P. Linna}
\email{Corresponding author: riku.linna@aalto.fi}
\affiliation{Department of Computer Science, Aalto University, P.O. Box 15400, FI-00076 Aalto, Finland}

\begin{abstract}

 We study translocation of semiflexible polymers driven by force $f_d$ inside a nanometer-scale pore using our three-dimensional Langevin dynamics model. We show that the translocation time $\tau$ increases with increasing bending rigidity $\kappa$. Similarly, the exponent $\beta$ for the scaling of $\tau$ with polymer length $N$, $\tau \sim N^\beta$, increases with increasing $\kappa$ as well as with increasing $f_d$. By comparing waiting times between semiflexible and fully flexible polymers we show that for realistic $f_d$ translocation dynamics is to a large extent, but not completely, determined by the polymer's elastic length measured in number of Kuhn segments $N_{\rm Kuhn}$. Unlike in driven translocation of flexible polymers, friction related to the polymer segment on the \textit{trans} side has a considerable effect on the resulting dynamics. This friction is intermittently reduced by buckling of the polymer segment in the vicinity of the pore opening on the {\it trans} side. We show that in the experimentally relevant regime, where viscosity is higher than in computer simulation models, the probability for this buckling increases with increasing $f_d$, giving rise to larger contribution to \textit{trans} side friction at small $f_d$. Similarly to flexible polymers, we find significant center of mass diffusion of the \textit{cis} side polymer segment. This speeds up translocation, which effect is larger for smaller $f_d$. However, this speed-up is smaller than the slowing down due to the \textit{trans} side friction. At large enough $N_{\rm Kuhn}$, the roles can be seen to be reversed and the  dynamics of flexible polymers be reached. However, for example, polymers used in translocation experiments of DNA, are elastically so short, that the finite-length dynamics outlined here applies.

\end{abstract}

\maketitle

\section{Introduction}\label{sec:intro}

Polymer translocation through a few nanometers wide pore has been studied extensively for two decades. The initial spur was given  by the seminal paper by Kasianowicz et. al. showing the potential of translocation in DNA sequencing ~\cite{Kasianowicz96}. Due to the potential applications and the originally missing theoretical framework, research has mostly concerned driven polymer translocation, where force is exerted on the polymer segment residing inside the pore~\cite{Sakaue_review}. This force drives the polymer from the initial \textit{cis} to the receiving \textit{trans} side. 

Most experiments on driven translocation are done on semiflexible polymers like DNA. In spite of this, all computer simulations in three dimensions until 2017 have been performed on translocation of fully flexible polymers. Also theoretically, fully flexible polymers have been of the highest interest. Very early Sakaue realized that the driven translocation dynamics is dominated by tension propagation in the polymer segment on the {\it cis} side~\cite{Sakaue07}. Early computer simulations confirmed this showing that during translocation, polymer segments on both sides of the membrane are driven out of equilibrium throughout the process~\cite{Lehtola09}. Sakaue formulated his tension propagation theory to completion, and an important addition was made by Rowghanian and Grosberg~\cite{Sakaue07,Sakaue11,Sakaue11Erratum,Rowghanian11,Sakaue_review}. It is now established that driven translocation dynamics of fully flexible polymers is determined by tension propagation on the \textit{cis} side, where monomers sequentially join the dragged segment of the polymer increasing the friction. Fluctuations, especially from the \textit{cis} side, have been shown to aid the process~\cite{Dubbeldam13,Suhonen14,Suhonen17}. The effect of monomer crowding on the {\it trans} side~\cite{Lehtola09} has been shown negligible~\cite{Suhonen14,Suhonen17}.

After the general properties of driven translocation are understood via flexible polymers, the attention in computer simulations is shifting toward translocation of semiflexible polymers. Invoking the concept of Kuhn length, a semi-flexible polymer in free space can be modeled at lower length-scale resolution as a flexible chain. This approach is applicable if the polymer is long enough compared to its persistence length. The approach obviously fails when polymers are confined. Such a case is a polymer ejecting from a viral capsid. Here semiflexibility changes the dynamics of translocation fundamentally in a way that cannot be accounted for by a simple change in the length scale~\cite{Linna17}. In driven translocation where polymer outside the pore is not confined it is an open question how well a change of length scale can account for changes brought in by polymer rigidity.

Driven translocation of semiflexible polymers has been studied by two-dimensional Langevin dynamics simulations~\cite{Adhikari13,Bhattacharya13}. Dynamics of translocating semiflexible polymers in two dimensions is expected to differ substantially from its three-dimensional counterpart. Three-dimensional simulations of semiflexible polymer translocation, accompanied by a simplifying model, have only recently been published~\cite{Sarabadani17}. Here the \textit{trans} side friction was found to have a significant contribution on translocation time. Furthermore, the buckling of the polymer segment on the \textit{trans} side was shown to reduce the friction and give it a characteristic dependence on the translocation coordinate $s$. However, the phenomenological characterization of buckling, which was found to be more probable for small than large pore force, leaves room for speculation.

Polymers used in experiments are relatively short compared to their persistence lengths. The longest polymers used in typical studies of Refs.~\cite{Storm05},~\cite{Fologea07},~and~\cite{Wanunu08} were only $323$, $77$, and $67$ Kuhn segments in length, respectively. This suggests that finite-size effects like the intermittently buckling semi-rigid polymer segment on the {\it trans} side, may have a determining role in real-world translocation. 

In what follows, we investigate driven semiflexible polymer translocation using our three-dimensional translocation model performing Langevin dynamics. On one hand, we set out to determine to what extent polymer rigidity can be accounted for by changing the relevant length scale. On the other hand, we will determine the effects of center of mass diffusion on the {\it cis} side and the elastic buckling on the {\it trans} side on the dynamics of driven semiflexible polymer translocation. We describe the used computational models in Section~\ref{sec:secx}, present and analyze the obtained results in Section~\ref{sec:results}, and make conclusions in Section~\ref{sec:conclusion}.

\section{The computational models}\label{sec:secx}
\begin{figure}[]
\includegraphics[width=0.80\linewidth]{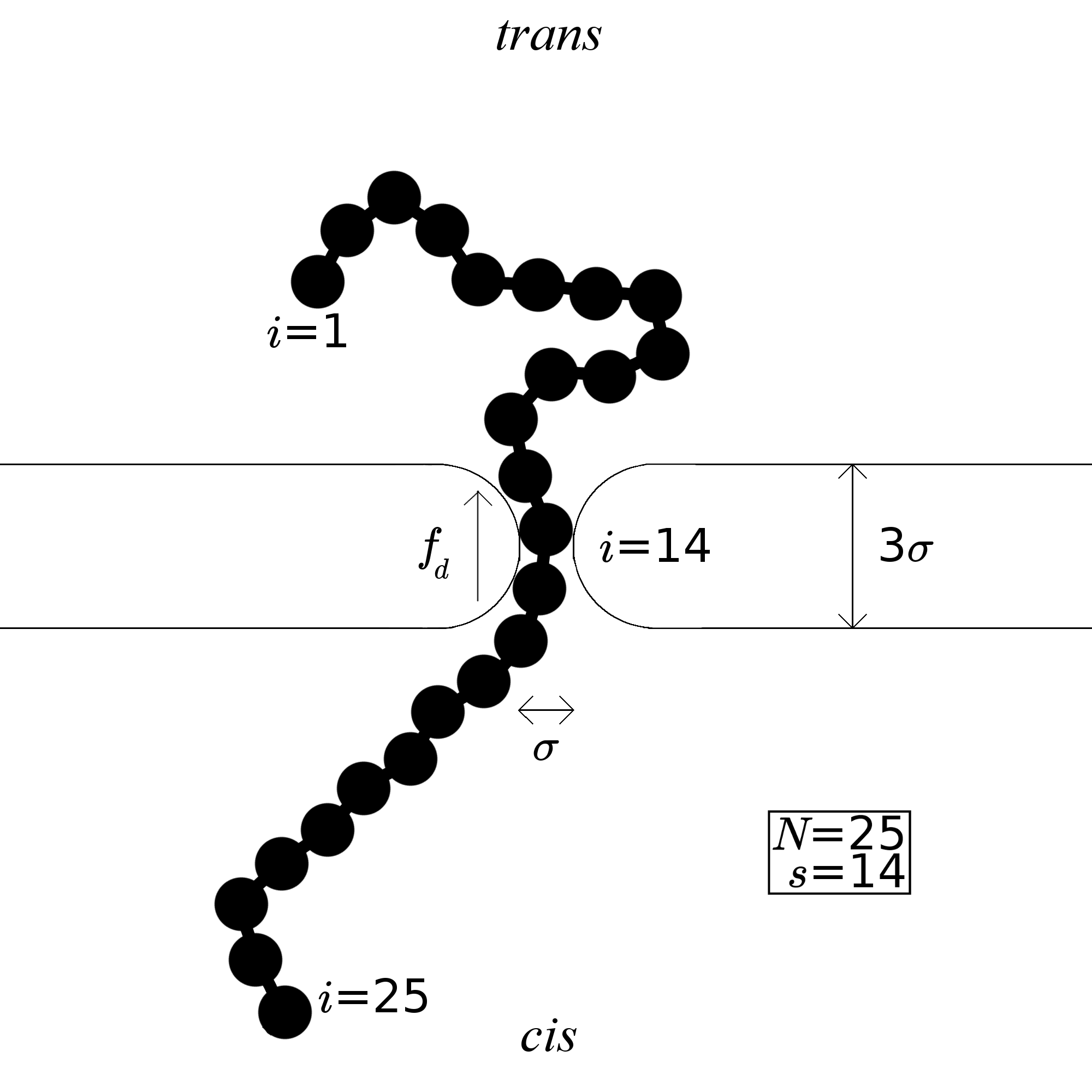}
\caption[]{A two dimensional cross section of the three-dimensional simulation setup. A polymer chain of length $N=25$ is driven through a small pore in a membrane from the \textit{cis} (bottom) to the \textit{trans} (top) side. Driving force $f_d$ is applied on the polymer segment in the pore ($f_d$ is divided among the beads currently occupying the pore). The membrane is $3 \sigma$ thick and the pore is $1 \sigma$ wide at the narrowest point. Polymer beads are indexed as $i=1$ for the first bead to translocate and $i=N$ for the last bead to translocate. In the figure, $14$ beads have crossed to the \textit{trans} side, so the translocation coordinate is $s=14$. The bead radius used in the figure was chosen for clarity. The interaction distance between a monomer and a wall is measured from the center of the bead. The figure is reprinted with permission from Ref.~\cite{Suhonen17} where aside from the bending potential, the simulation setup is the same.}
\label{fig:simusetup}
\end{figure}

Here we describe the computational model for our three-dimensional Langevin dynamics simulations. Except for the added bending potential, the present model is similar to what we have used in~Refs.~\cite{Suhonen17}~and~\cite{Suhonen14}. In addition, the pore model is slightly different from the one used in Ref.~\cite{Suhonen14}. The simulation setup is depicted in Fig.~\ref{fig:simusetup}.

\subsection{Polymer model}
We model the polymer as a worm-like chain (WLC) consisting of a standard bead spring model combined with a bending potential. Any two beads in the model repel each other via a truncated and shifted Lennard-Jones (LJ) potential given as
\begin{align}\label{LJ}
U_{\rm LJ} = \left\lbrace \begin{array}{ccl}
    4 \epsilon \left[\left(\frac{\sigma}{r}\right)^{12}-\left(\frac{\sigma}{r}\right)^{6} + \frac{1}{4}\right] &,& r \leq 2^{1/6} \sigma
\\
0&,& r > 2^{1/6} \sigma
\end{array},
\right.
\end{align}
where $r$ is the distance between the two beads. $\epsilon=1$ and $\sigma=1$ define the units of energy and length in the model.

Two adjacent beads in the chain are connected via finitely extensible nonlinear elastic (FENE) potential given as 
\begin{align}\label{FENE}
\begin{array}{ccl}
U_{\rm F} = -\frac{K}{2}R^2 \ln{(1-\frac{r^2}{R^2})} &, & r<R.
\end{array}
\end{align}
$R=1.5 \sigma$ is the maximum elongation of the bond, and $K=30 \sigma$ is the strength of the interaction.

The bending potential energy of the whole polymer is defined as
\begin{align}\label{FENE}
\begin{array}{ccl}
U_{\rm bend} = -\kappa \sum\limits_{i} (\boldsymbol{r}_{i+1}-\boldsymbol{r}_{i}) \cdot (\boldsymbol{r}_{i}-\boldsymbol{r}_{i-1})
\end{array}
\end{align}
Here, the sum runs over all beads $i$ in the polymer. $\boldsymbol{r}_{i}$ is the position vector of bead $i$ and $\kappa$ is the bending rigidity parameter. For $\kappa=0$, we have estimated from simulations that the persistence length $\lambda_p \approx 0.9$. This exceeds $\lambda_p= \frac{1}{2}$ (half bond length) due to excluded volume interactions. For higher $\kappa$ values, $\lambda_p \approx \kappa$. Polymer beads are indexed so that bead of index $i=1$ translocates first and bead of index $i=N$ translocates last.

\subsection{Simulation algorithm}

We use Ermak's implementation of Langevin dynamics~\cite{Ermak80,Allen}. Movement of all beads is determined by the Langevin equation
\begin{align}\label{eq:langevin}
\dot{\textbf{p}}_i=-\xi\textbf{p}_i+\pmb{\eta}_i(t)+\textbf{f}_i(\textbf{r}_i),
\end{align}
where $\textbf{p}_i$ is the momentum, $\xi=0.5$ is the friction coefficient, $\pmb{\eta}_i$ is the Langevin random force, and $\textbf{f}_i$ is the resultant of all other forces affecting the particle. For $\pmb{\eta}_i(t)$ it holds that $\langle \pmb{\eta}_i(t) \rangle=0$ and $\langle \pmb{\eta}_i(t) \pmb{\eta}(t')\rangle=2 \xi k T m \delta(t-t')$, where $\delta(t)$ is the Dirac delta function, $k$ is the Boltzmann constant, $T$ is the temperature, and $m$ is the mass of a bead. We set $kT=1$ and $m=16$. To integrate the equations of motion, we use the velocity Verlet algorithm~\cite{vanGunsteren77}. For the integration, we set the time step $\delta t=0.001$ for translocation and $\delta t=0.025$ for equilibration prior to translocation.

\subsection{Pore and membrane models}
The geometrical aspects of the simulation setup are depicted in Fig.~\ref{fig:simusetup}. We model the membrane as two impermeable planes at $r_z=-1.5 \sigma$ and $r_z=1.5 \sigma$. Points for which $r_z<-1.5 \sigma$ are on the \textit{cis} side and points for which $r_z>1.5 \sigma$ are on the \textit{trans} side. The pore is modeled as an opening around $r_x=r_y=0$. The opening has the shape of a hole in the middle of a torus. At the narrowest point the diameter of the pore is $\sigma$. The length of the pore is $3 \sigma$. The geometries are simulated using constructive solid geometry technique~\cite{Wyvill85} applied to molecular dynamics simulations~\cite{Piili15,Piili17,JoonasPhd}. The membrane and pore geometries apply slip boundary conditions to beads colliding with the geometry. The translocation is driven by force $f_d$ toward positive $z$-axis. $f_d$ is divided among the beads inside the pore.

\subsection{The relation to real DNA molecules}
Reduced units are used in our simulations ~\cite{Allen}. Here, we present a non-comprehensive mapping between our simulation space and the real world. We take our model polymers of $\kappa=20$ to represent double stranded DNA (dsDNA). The persistence length of dsDNA is around $50$ nm, corresponding to $150$ base pairs (bp)~\cite{Smith92,Manning06,Geggier11}. For $\kappa=20$, $\lambda_p \approx 20 b$, where $b \approx \sigma$ is the bond length in our simulations. Now $\sigma = \frac{50 {\rm ~nm}}{20} \approx  2.5 {\rm ~nm}$. The thickness of the membrane is then $3 \sigma=7.5 {\rm ~nm}$, which is comparable with the lipid bilayer thickness of $\approx 4 {\rm ~nm}$~\cite{Mitra04}. The shortest and longest polymers used in this article are $N=25$ and $N=400$. They correspond to the lengths $63 {\rm ~nm}$ and $1 {\rm ~\mu m}$ ($190 {\rm ~bp}$ and $3000 {\rm ~bp}$), respectively. These lengths are comparable to dsDNA chains of $30 {\rm ~bp}$ to $20000 {\rm ~bp}$ used in experiment of Ref~\cite{Carson14}. Each simulation monomer corresponds to $\frac{150 {\rm ~bp}}{20}=7.5 {\rm ~bp} = 15$ nucleotides. The above mapping is only for the length scale. Relations for mass, energy and time scales can also be calculated. As the mappings are not exact, we omit further treatment and refer the  interested reader to~\cite{Suhonen17} where we discuss the mapping in more detail in the case of flexible chains.

\section{Results}\label{sec:results}
Here, we present the central results of our computer simulations on driven polymer translocation and discuss the characteristics pertinent to the bending rigidity of the simulated semi-flexible polymers. The results are generally ensemble averages over about 500 simulations for each parameter set. For some parameters, more simulations were used for obtaining better signal-to-noise ratio. In the initial conformations there are two beads on the \textit{trans} side and three beads inside the pore. The rest of the beads are on the \textit{cis} side. A bead is considered as translocated when it reaches the middle of the pore. Hence, at the start three beads are already considered translocated. These beads are subtracted from results involving the length of the polymer. Initially, before letting each polymer translocate it is equilibrated until its radius of gyration has converged to a stable value.

In Sec.~\ref{sec:waiting_times}, we report the measured waiting times and show how translocation dynamics changes due to bending rigidity. In Sec.~\ref{sec:kappamapping}, we show that the effects of bending rigidity are partially explained by polymer length in terms of Kuhn segments. In Sec.~\ref{sec:diffusion}, we examine diffusive effects on the \textit{cis} side. In Sec.~\ref{sec:buckling}, we study buckling of the semi-flexible polymer segment on the \textit{trans} side. Lastly, in Sec.~\ref{sec:scaling}, we discuss the effect of bending rigidity on the scaling exponents of driven polymer translocation.

\subsection{Waiting times $t_w$}\label{sec:waiting_times}

\begin{figure}[]
\includegraphics[width=0.49\linewidth]{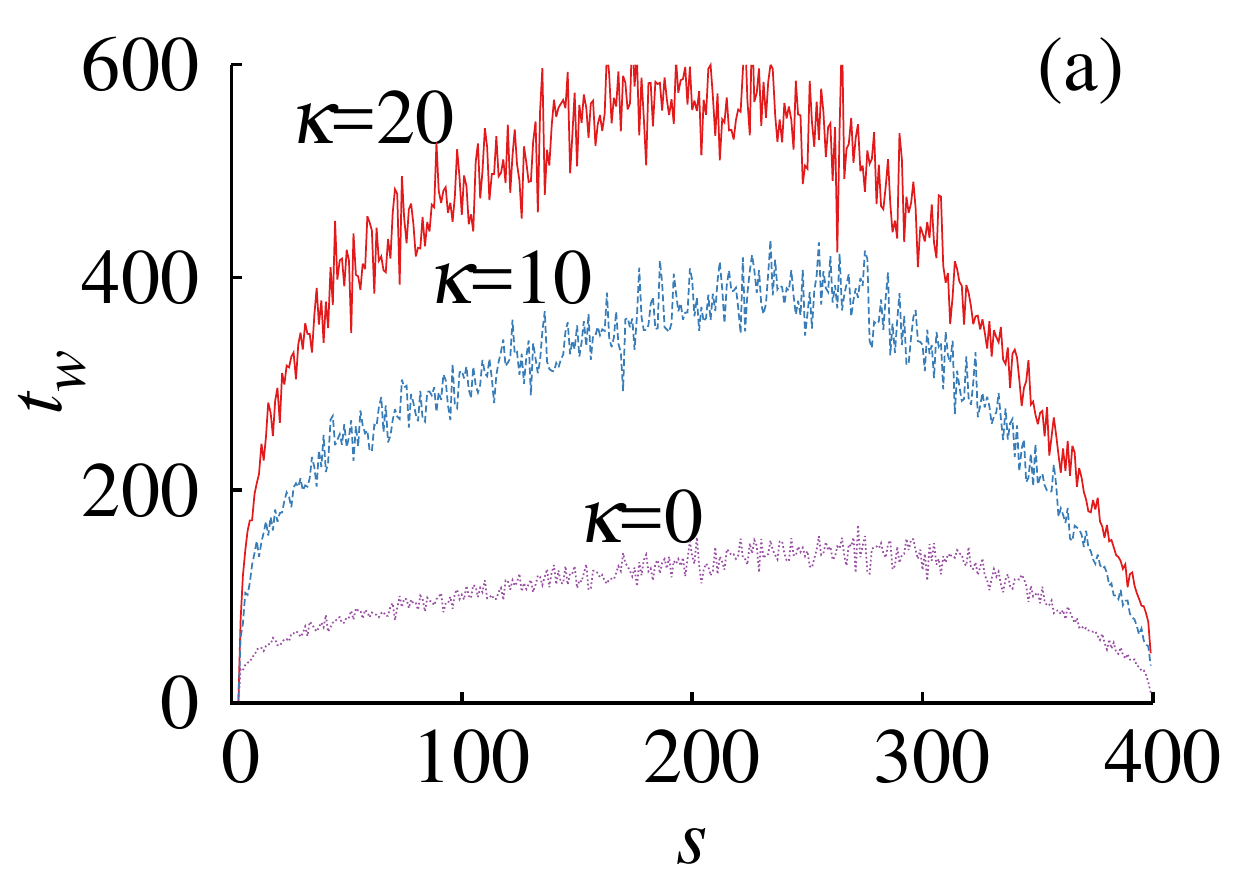}
\includegraphics[width=0.49\linewidth]{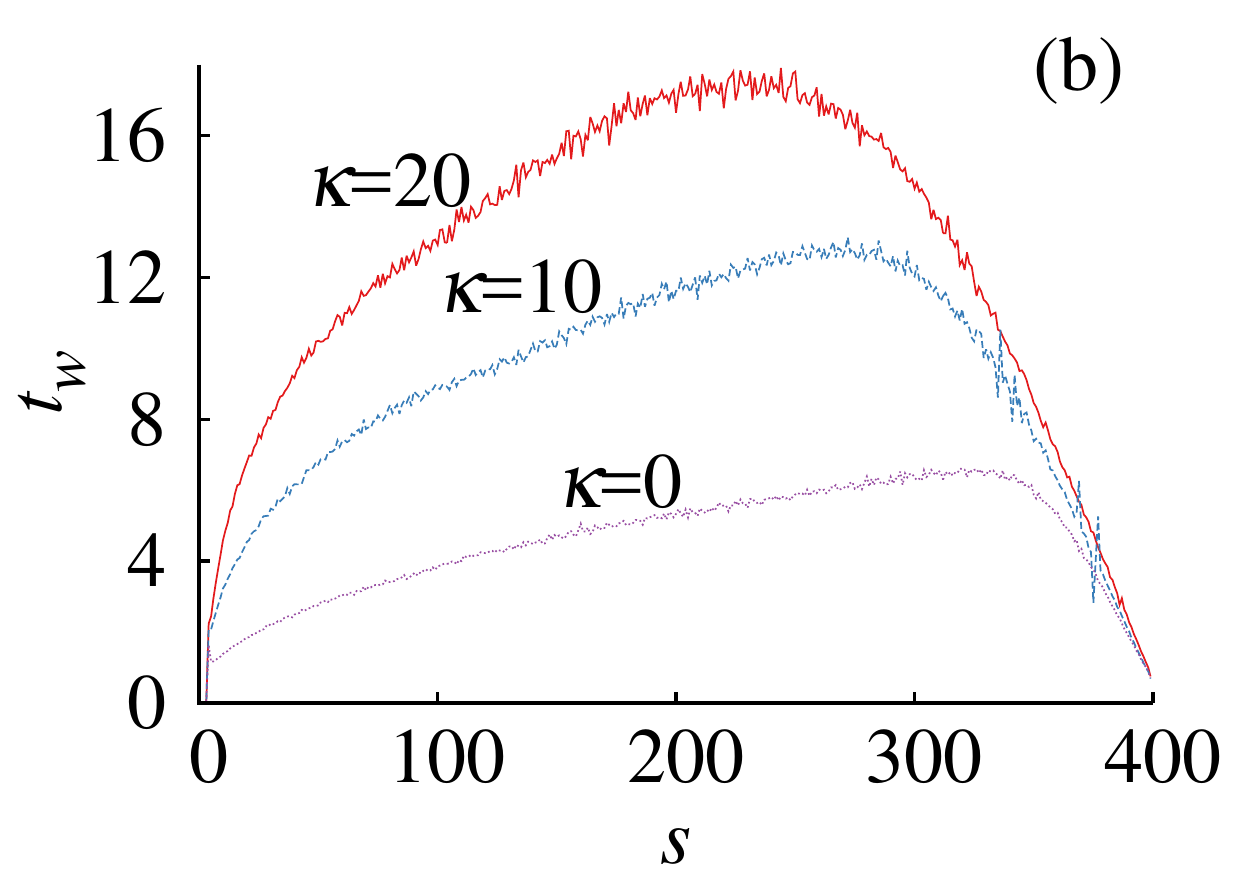}
\caption[]{(Color online) Waiting times $t_w$ vs. number of translocated monomers $s$. Polymer length $N=400$. Rigidity from bottom to top: $\kappa=0$, $10$, and $20$. Pore force (a) $f_d=2$ (b) $f_d=64$.}
\label{fig:k20k10k0_N400_fpwt}
\end{figure}

We define waiting time $t_w(s)$ as the interval between instants that polymer beads (monomers) $s$ and $s-1$ enter the \textit{trans} side for the first time.  $t_w$ for polymers of length $N=400$ and  rigidity $\kappa=0$, $10$, and $20$ are shown in Fig.~\ref{fig:k20k10k0_N400_fpwt}~(a) for $f_d=2$ and in Fig.~\ref{fig:k20k10k0_N400_fpwt}~(b) for  $f_d=64$. Polymer bending rigidity is seen to slow down translocation. Moreover, increasing $\kappa$ shifts the peak of the $t_w$ curve to smaller $s$, which means that tension propagation ends and retraction (also called post propagation) starts earlier. In addition, for $f_d=2$ the peaks are more rounded and occur at smaller $s$ than for $f_d=64$. This is caused by diffusion of the polymer conformation on the {\it cis} side which we previously found out to be responsible for the same characteristic for flexible polymers~\cite{Suhonen17}. The effect is stronger for smaller $f_d$, but can be seen in the strong $f_d$ regime to which all our simulations belong to.

\begin{figure}[]
\includegraphics[width=0.49\linewidth]{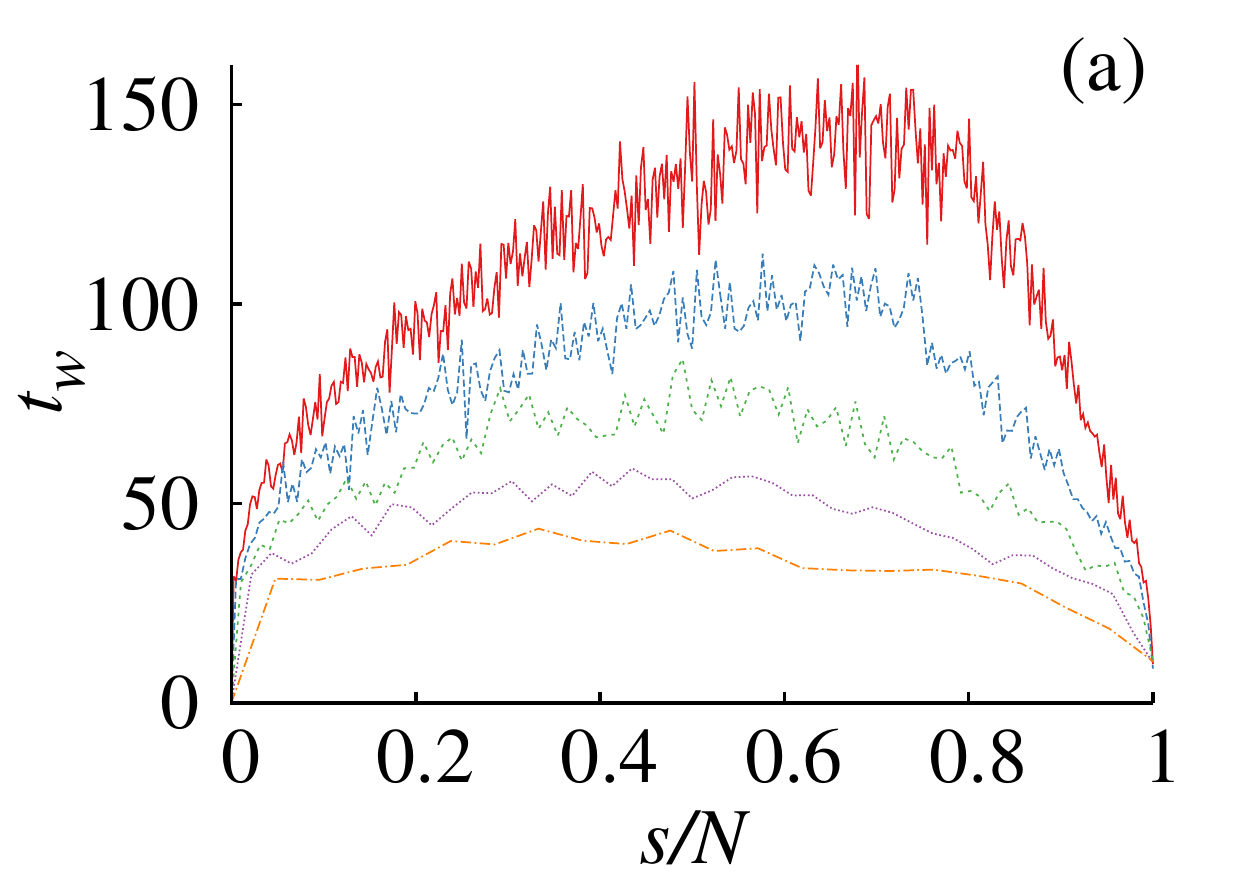}
\includegraphics[width=0.49\linewidth]{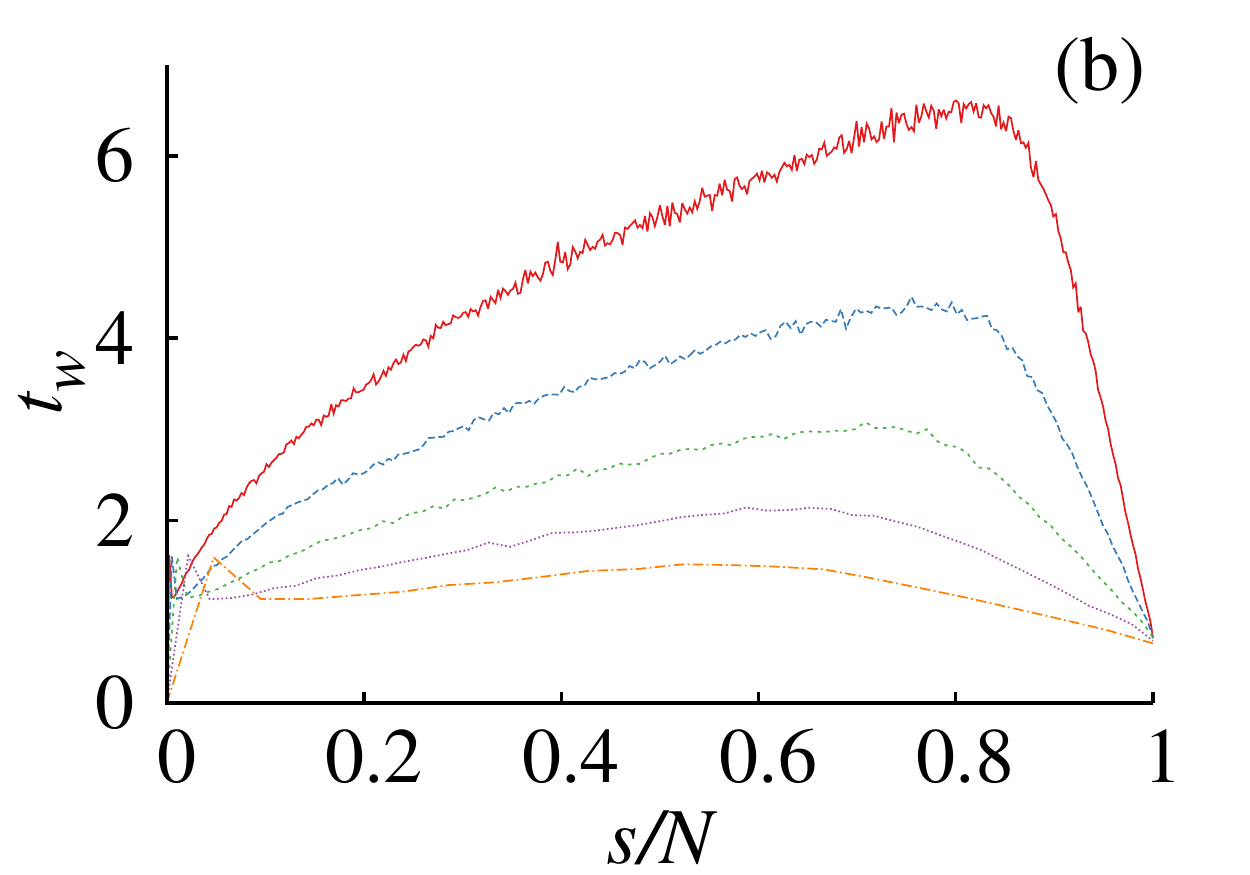}
\caption[]{(Color online) Waiting times $t_w$ vs. normalized reaction coordinate $s/N$ for fully flexible polymers ($\kappa=0$). From top to bottom $N=400$, $200$, $100$, $50$, and $25$. Pore force (a) $f_d=2$ and (b) $f_d=64$.}
\label{fig:k0_N25to400_fpwt}
\end{figure}

Characteristics for different $N$ can be compared by plotting $t_w$ as a function of normalized reaction coordinate $s/N$. Figs.~\ref{fig:k0_N25to400_fpwt}~(a)~and~(b) show $t_w(s/N)$ for fully flexible ($\kappa=0$) polymers of lengths $N=25$, $50$, $100$, $200$, and $400$ for $f_d=2$ and $f_d=64$. For shorter polymers $t_w$ peaks occur at smaller $s/N$.

\begin{figure*}
\includegraphics[width=0.3\linewidth]{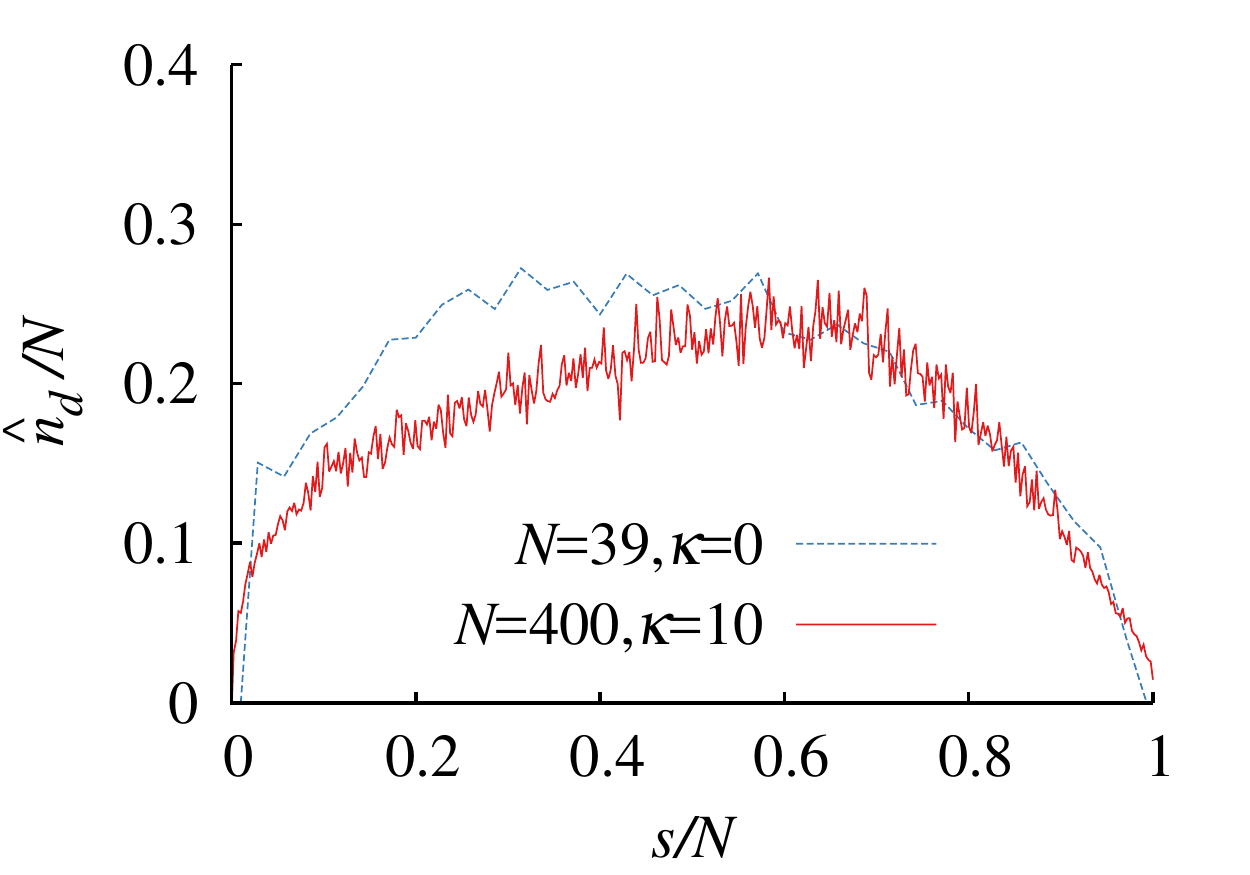}
\includegraphics[width=0.3\linewidth]{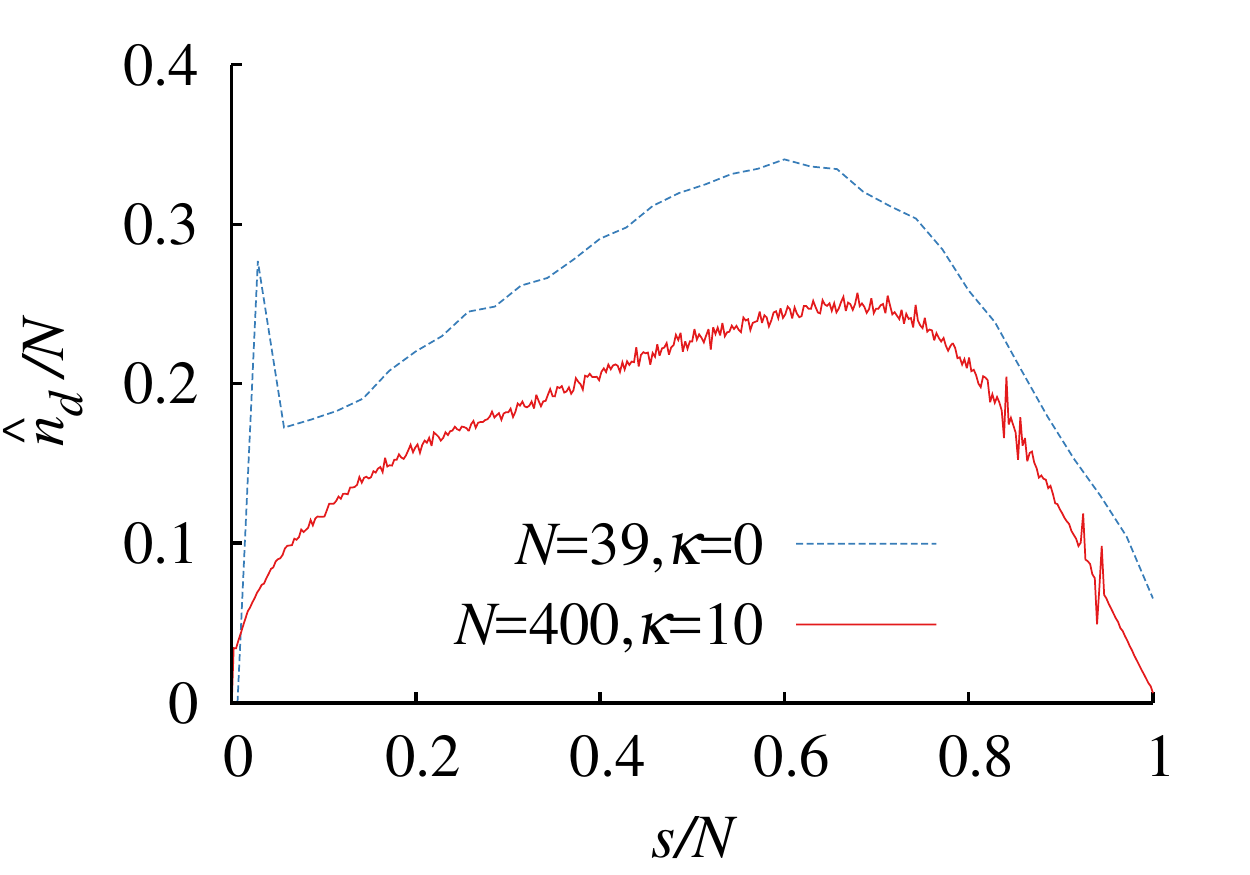}
\includegraphics[width=0.3\linewidth]{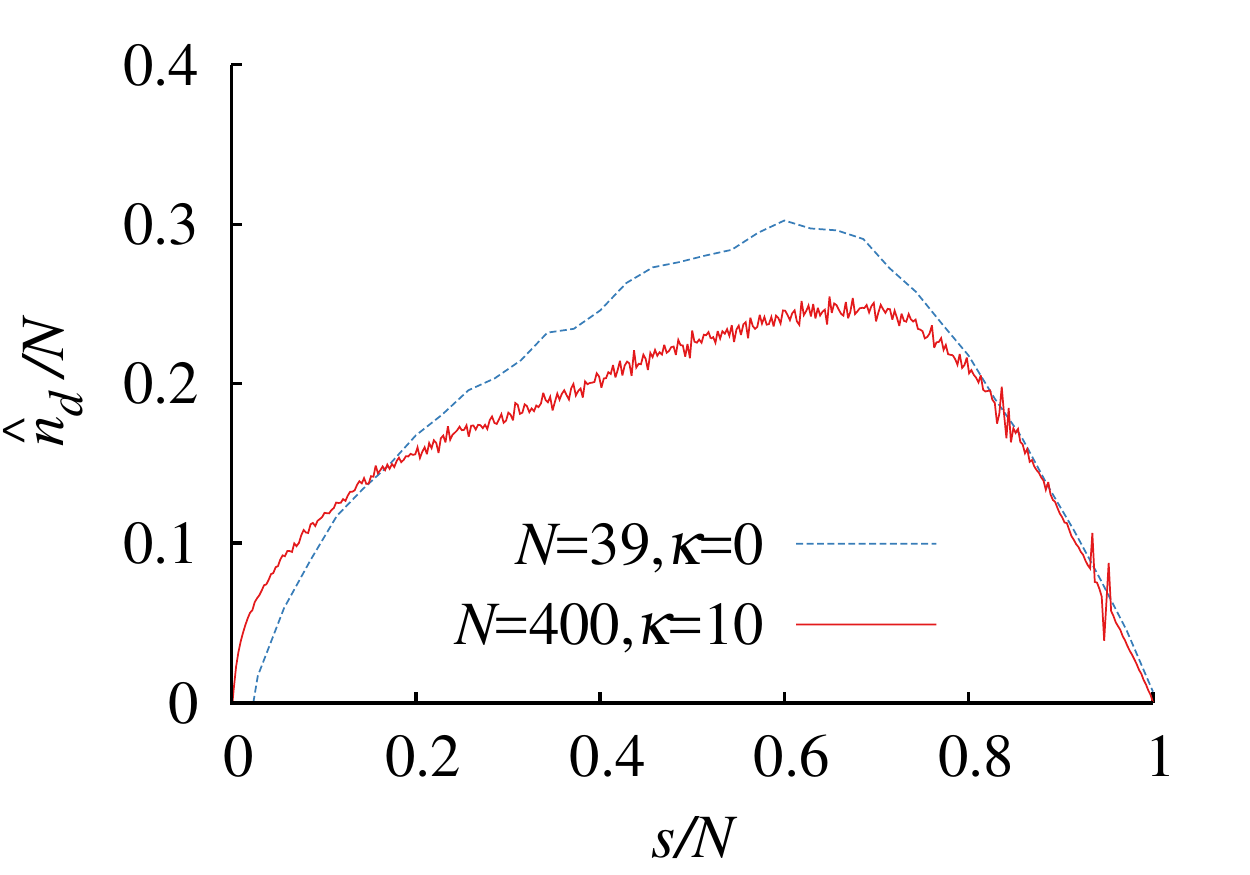}\\
\includegraphics[width=0.3\linewidth]{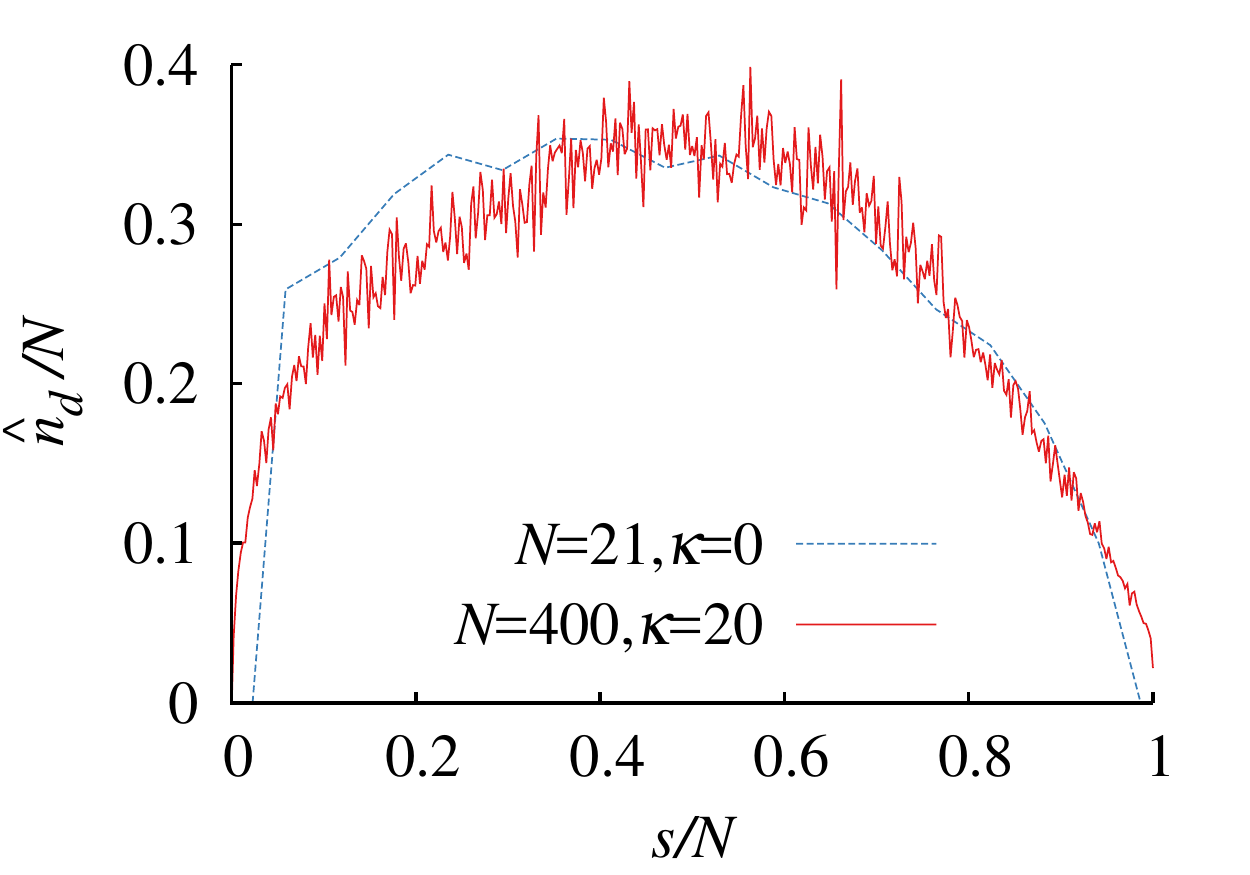}
\includegraphics[width=0.3\linewidth]{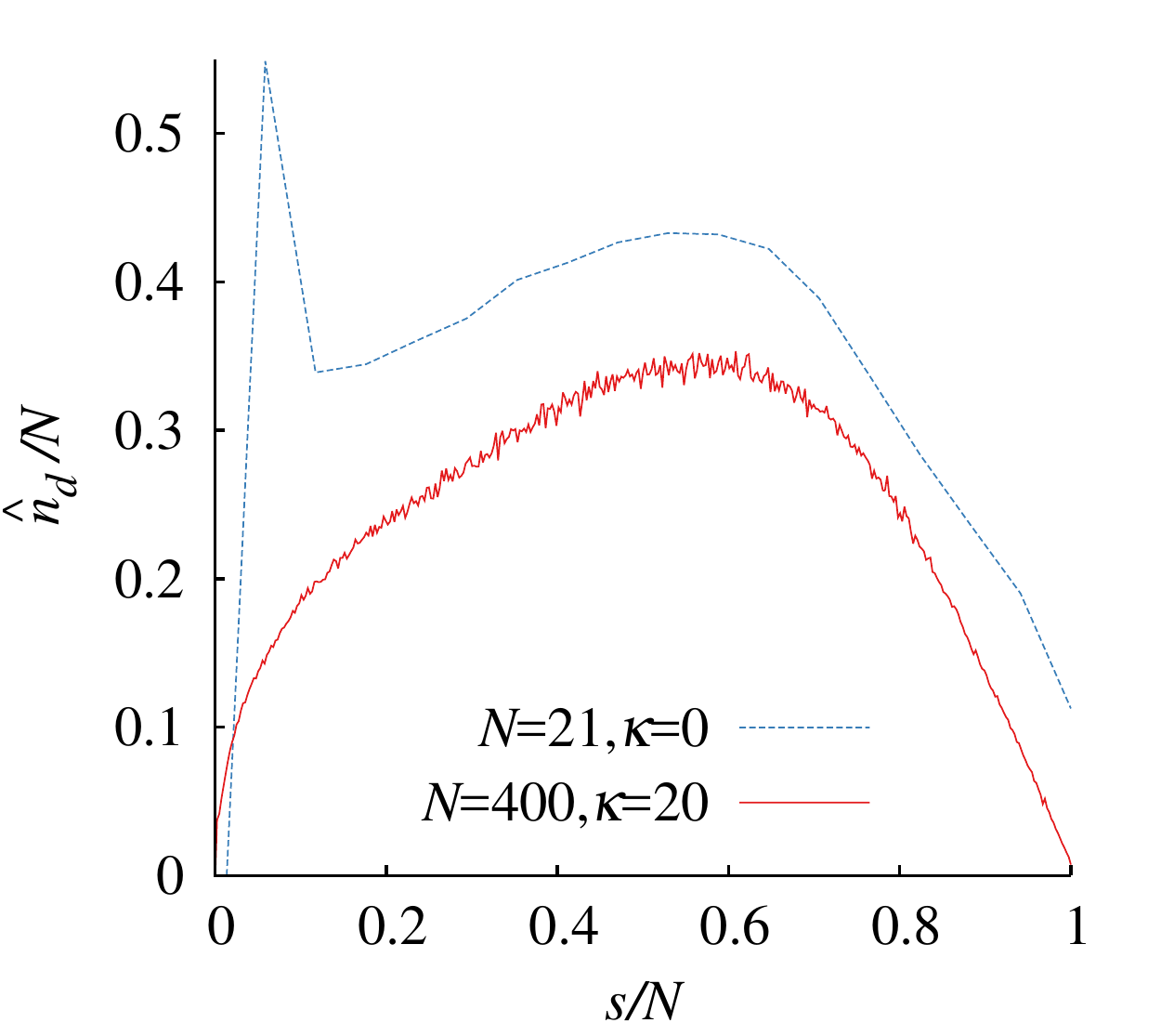}
\includegraphics[width=0.3\linewidth]{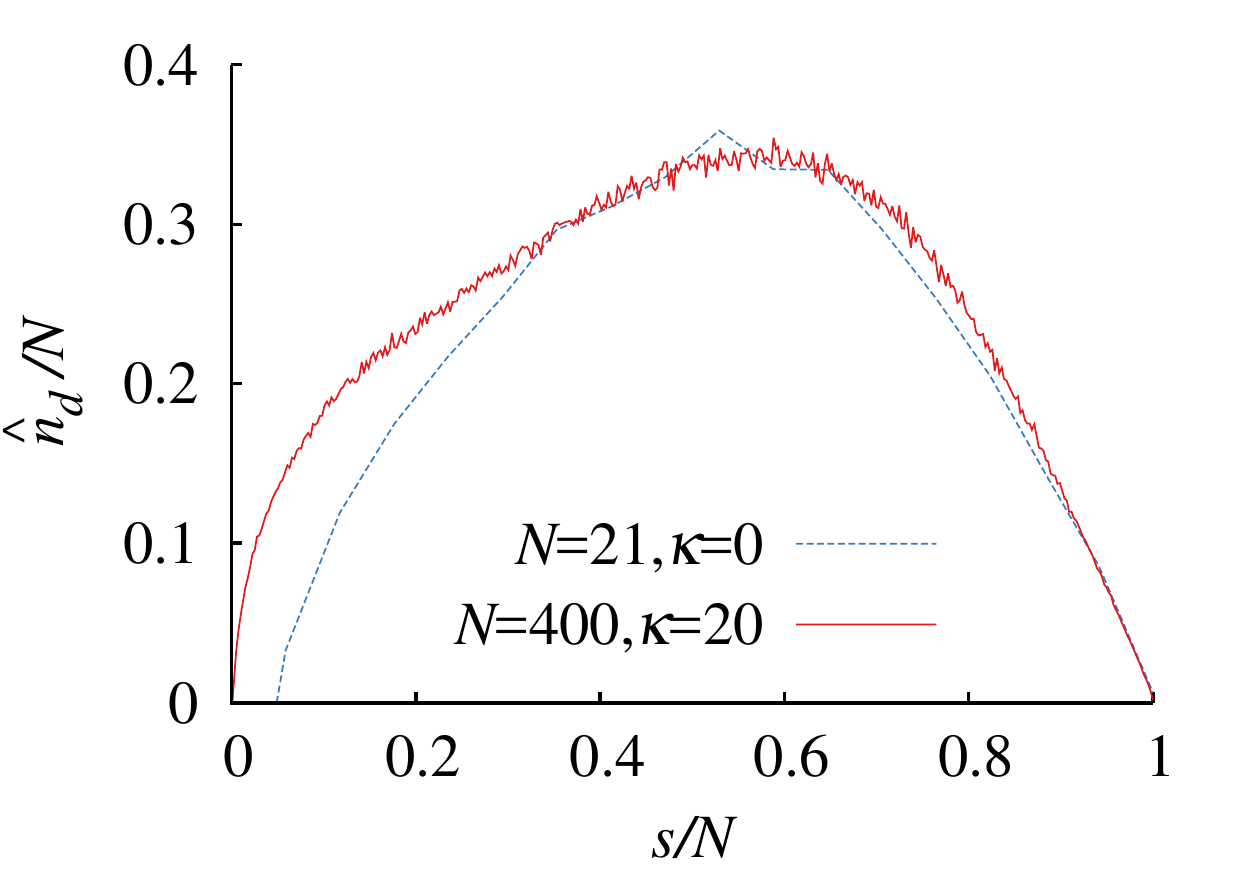}
\caption[]{(Color online) The fraction of the total length of polymer under drag, $\hat{n}_d/N$, as a function of normalized translocation coordinate $s/N$. Polymers of comparable lengths in number of Kuhn segments are shown in each plot (see text). The first row: The first set of polymers has $N = 39$, $\kappa = 0$, and $N_{\rm Kuhn} = 20$. The second set of polymers has $N = 400$, $\kappa = 10$, and $N_{\rm Kuhn} \approx 19.9$. The second row: The first set of polymers has $N = 21$, $\kappa = 0$, and $N_{\rm Kuhn}  10$. The second set of polymers has $N = 400$, $\kappa = 20$, and $N_{\rm Kuhn} \approx 9.9$. The first column: $\xi = 0.5$, $f_d=2$. The second column: $\xi = 0.5$, $f_d=64$. The third column: $\xi = 100$, $f_d=64$.}
\label{fig:kappaNScaled_tw}
\end{figure*}

The shift in the peaks of both Figs.~\ref{fig:k20k10k0_N400_fpwt}~and~\ref{fig:k0_N25to400_fpwt} can be explained through the Kuhn length. Increasing $\kappa$ increases the Kuhn length, a statistical measure for polymer rigidity defined as the ratio of polymer's average squared radius and the maximum end-to-end distance, $l_{\rm Kuhn} = \langle R^2 \rangle/R_{\rm max}$~\cite{DoiEdwards}. The Kuhn length can be expressed as $l_{\rm Kuhn} = n_{\rm Kuhn}b$, where $n_{\rm Kuhn}$ is the number of beads in a segment of length $l_{\rm Kuhn}$ and $b$ is the equilibrium length of a polymer bond. A reasonable candidate for the effective length of a polymer, to take into account its rigidity, is the number of Kuhn segments $N_{\rm Kuhn} = N/n_{\rm Kuhn}$. Either increasing $\kappa$ for constant $N$ or reducing $N$ and keeping $\kappa$ constant leads to shorter polymers in terms of $N_{\rm Kuhn}$. In accordance, maxima in $t_w$ are seen to shift toward smaller $s$ when increasing $\kappa$, Fig.~\ref{fig:k20k10k0_N400_fpwt}, and smaller $s/N$ when reducing $N$, Fig.~\ref{fig:k0_N25to400_fpwt}. This trend is conceivable, since in the extreme case of increasing $\kappa$ sufficiently so that $\lambda_p > Nb$, the peak in $t_w$ disappears altogether, and $t_w(s)$ becomes flat.

\subsection{Comparison of polymers with different $N$ and $\kappa$}\label{sec:kappamapping}

In free space and close to thermal equilibrium, semiflexible polymers of persistence length $\lambda_p$ can be modeled as fully flexible chains of bond length $b=2 \lambda_p$ by changing the degree of coarse-graining~\cite{Grosberg_book94,Rubinstein16}. In view of the findings of Sec.~\ref{sec:waiting_times}, it needs to be assessed to what extent bending rigidity can be taken into account in the context of driven translocation by merely adjusting the relevant length scale. We do this by estimating from simulations  the ratio of the length of the polymer segment causing friction to the total polymer length $n_d/N$. $n_d = n_d^c+n_d^t+n_d^p$, where $n_d^c$ is the number of beads under drag on the \textit{cis} side, $n_d^t$ is the number of beads currently being pushed on the \textit{trans} side, and $n_d^p$ is the number of beads causing friction inside the pore.

In order to discuss the effective friction during translocation, we transform the measured waiting times to number of monomers in drag. This transformation is done conceptually in the overdamped limit and neglecting fluctuations, where we have $\textbf{f}_i(\textbf{r}_i)=\xi\textbf{p}_i$, see Eq.(\ref{eq:langevin}). Then $f_d=n_d(s) \xi m v(s)$, where $v(s)$ is taken as average velocity of the $n_d$ beads at $s$. Waiting times $t_w(s)$ can be given in terms of velocity $v(s)$ and bond length $b$ as $t_w(s)=b/v(s)$. Since $b \approx \sigma$, we obtain $n_d(s)=\frac{t_w(s) f_d}{\xi m \sigma}$. The pore is $3 \sigma$ long, so approximately $n_{d}^p=3$ beads reside inside the pore throughout the translocation. To reduce the effect caused by the beads in the pore, which is relatively stronger for shorter polymers, we define $\hat{n}_d=n_d-n_{d}^p$. Since this transformation would be exact only in the overdamped limit with insignificant fluctuations, all effects, including those from diffusion, are included in $n_d$. This transformation of $t_w$ to $n_d$ is introduced in order to make the following discussion more clear.

Figure~\ref{fig:kappaNScaled_tw} shows $\hat{n}_d/N$ curves of a semiflexible and fully flexible polymers for $N$ and $\kappa$ chosen such that the polymer lengths in Kuhn segments $N_{\rm Kuhn}$ are comparable. These lengths are calculated as $N_{\rm Kuhn} = (N-3)/(2 \cdot \lambda_p)$. Above, three beads are subtracted to account for the beads that are already translocated in the initial conformation of our translocation simulations. Semiflexible polymers of two different rigidity are compared with fully flexible polymers. On the first row $\kappa = 10$ and  on the second row $\kappa = 20$. The first two columns are for $f_d=2$ and $64$ using friction $\xi = 0.5$. On the third column $f_d=64$ and $\xi = 100$. The persistence length $\lambda_p \approx 0.9$, $10$, and $20$ for $\kappa = 0$, $10$, and $20$, respectively.

$n_d/N$ for flexible and semiflexible polymers of comparable $N_{\rm Kuhn}$ are in general agreement for $f_d = 2$ (first column of Fig.~\ref{fig:kappaNScaled_tw}). $n_d/N$ for $s/N \lesssim 0.6$ is greater for fully flexible chains. These chains are very short in order to make them comparable to the semiflexible chains. The polymers are of thickness $\sigma$. Therefore, excluded volume interactions show more prominently for short polymers, and consequently the extension of polymers' initial conformation normalized with the polymer length is greater for short polymers. As translocation proceeds and the polymer extends, the excluded volume interactions decrease and vanish. In agreement with this $n_d/N$ for short fully flexible polymers and semiflexible polymers are seen to align  for $s/N \gtrsim 0.6$.

Increasing bias to  $f_d = 64$ (second column of Fig.~\ref{fig:kappaNScaled_tw}) the agreement between flexible and semiflexible gets poorer. The short fully flexible chain shows a large (inertial) transient, which is to be expected at such large bias. For large $f_d$, the pore friction is enhanced due to geometric effects local to the pore entrance~\cite{Suhonen14,Suhonen17}. This effect shows more prominently for short polymers, since under identical viscous conditions friction outside the pore is larger for longer polymers. For any $s/N$ the same relative fraction of polymers are tensed and under drag for polymers of different $N$. The frictional contribution from the pore leads to overestimation of ${n}_d/N$. This overestimation is larger for shorter polymers. Increasing viscosity outside the pore decreases the effect from the pore. This is seen on the rightmost column in Fig.~\ref{fig:kappaNScaled_tw} where $\hat{n}_d/N$ are shown for $\xi = 100$.

It is in place to note again that in order to relate the length scale to polymer rigidity, we used the assumption of no fluctuations and strongly damped dynamics. Hence, all dynamics is reflected in the calculated friction or the number of beads in drag $n_d$. Clearly, for polymers of lengths used here there is a contribution on semiflexible polymers that cannot be accounted for by a simple length-scale transformation based on rigidity. To better determine the remaining contribution we next look at the contribution of diffusion in the next section.

\subsection{Diffusion on the \textit{cis} side}\label{sec:diffusion}

Previously, we observed clear $f_d$ dependent changes in the dynamics of driven translocation of fully flexible polymers. We showed these changes to be largely due to the center of mass diffusion of the polymer on the \textit{cis} side~\cite{Suhonen17}. Here, we show this center of mass diffusion to have a clear effect also on driven translocation dynamics of semiflexible polymers.

\begin{figure}[]
\begin{centering}
\includegraphics[width=0.80\linewidth]{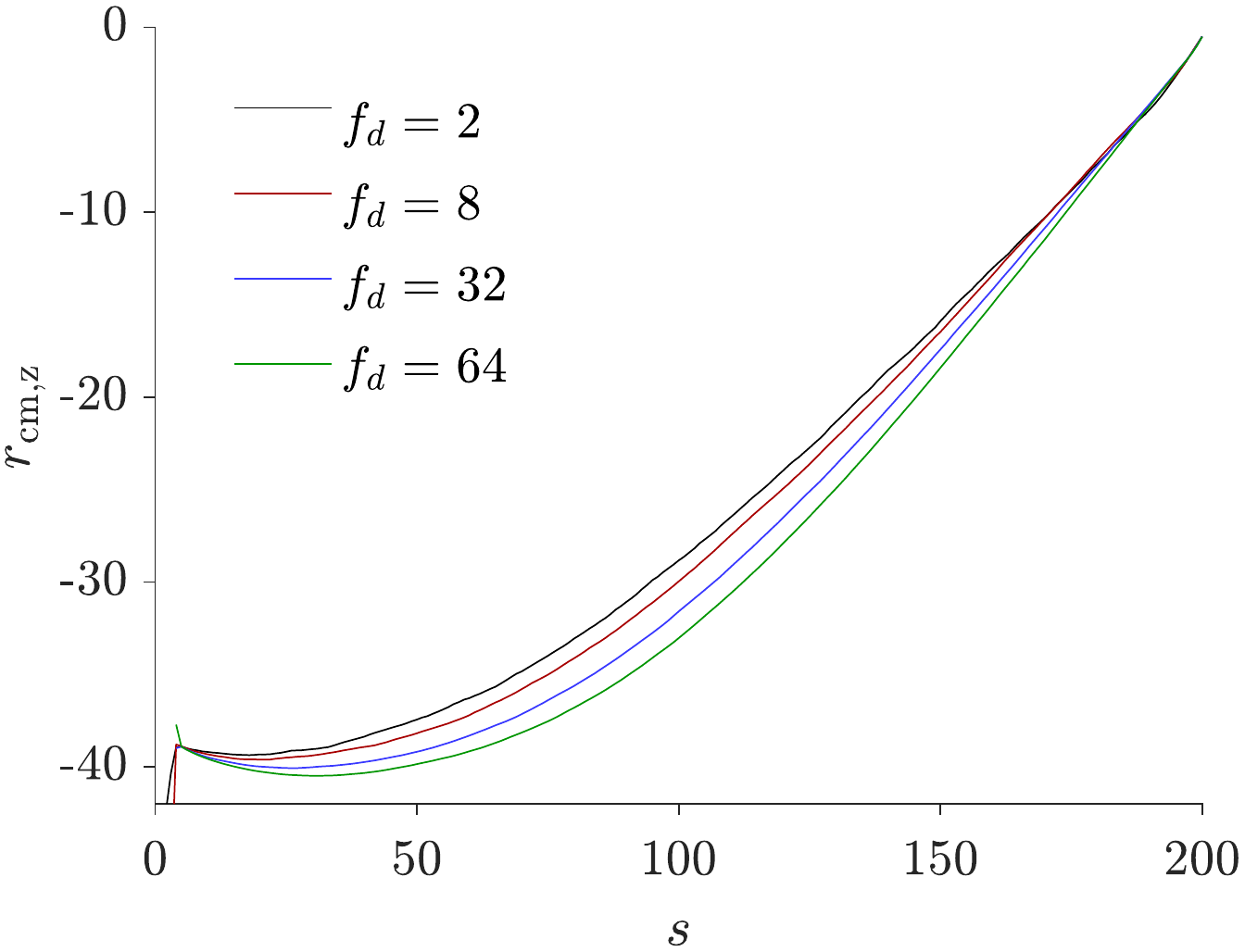}
\caption[]{(Color online) z-component of the center of mass position $r_{\rm cm,z}$ of the \textit{cis} side polymer segment. Measured for polymers of $N=200$ and $\kappa=20$. From top to bottom: $f_d=2$, $8$, $32$, and $64$. Pore center is located at $r_{\rm cm,z}=0$.}
\label{fig:CMCis_wlc2}
\end{centering}
\end{figure}

Fig.~\ref{fig:CMCis_wlc2} shows the $z$-coordinate of the center of mass position of the \textit{cis} side polymer segment for $N=200$ and $\kappa=20$ polymers for various $f_d$. The position of the pore is at $r_z=0$. It is seen that for small $f_d$ the polymer's center of mass approaches the pore faster in $s$. i.e. faster with respect to the translocation time $\tau$. Without fluctuations, the center of mass movement as a function of $s$ would be similar regardless of $f_d$~\cite{Suhonen17}. Hence, the characteristics of Fig.~\ref{fig:CMCis_wlc2} results from center of mass diffusion towards the pore. This confirms what was anticipated from indirect observations in Sections~\ref{sec:waiting_times}~and~\ref{sec:kappamapping}.

In summary, the center of mass diffusion has a clear effect on both flexible and semiflexible polymer translocation. Making the length scales based on rigidity comparable in Sec.~\ref{sec:kappamapping} resulted in fair alignment of the driven translocation dynamics of flexible and semiflexible polymers for moderate $f_d$. As described, the diffusive effects were lumped into effective friction. The alignment would imply that taking into account the effective polymer length determined by polymer rigidity transforms also diffusive effects satisfactorily for moderate $f_d$.

\subsection{\textit{Trans} side friction and buckling}\label{sec:buckling}

In contrast to fully flexible polymers, due to increased polymer chain rigidity, \textit{trans} side friction can potentially be a significant factor in semiflexible chains of small $N_{\rm Kuhn}$. In a recent study the effect of \textit{trans} side friction and chain buckling characteristics were reported for one set of parameters and used in order to show agreement between a model and experimental results on DNA translocation ~\cite{Sarabadani17}. Here, we show that these characteristics are highly dependent, for example, on viscosity and pore bias $f_d$.

\begin{figure}[]
\includegraphics[width=1.00\linewidth]{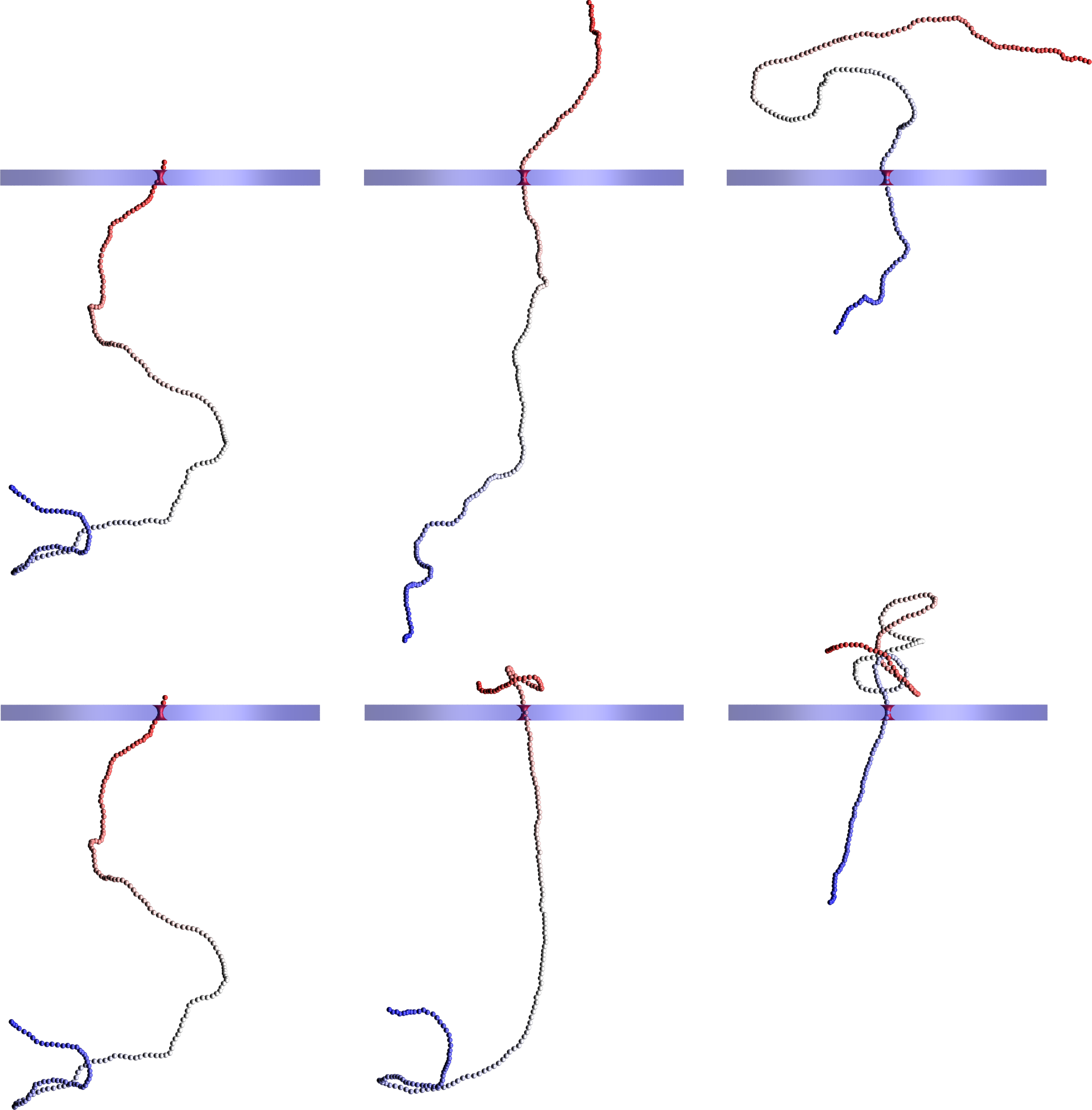}
\caption[]{(Color online) Snapshots from simulations for $\kappa=20$ and $N=200$. Top row: $f_d=2$. Bottom row: $f_d=64$. From left to right snapshots are for translocation coordinates $s=4$, $50$, and $150$. For smaller $f_d$ the polymer extends much further on the \textit{trans} side. Opposite is true for the tensed segment on the \textit{cis} side where center of mass diffusion alleviates the tension for lower $f_d$.} 
\label{fig:snapshots_f2_and_f64_from_run1}
\end{figure}

Fig.~\ref{fig:snapshots_f2_and_f64_from_run1} shows typical snapshots from our simulations of polymers of $N=200$, $\kappa=20$, and  $f_d=2$ and $64$. Both of the presented simulations start from the same configuration. As the polymer translocates, the \textit{trans} side segment first stays relatively straight and then bends and buckles. For higher $f_d$, there is clearly a higher tendency for this buckling. So much so that the initially straight segment is barely visible for $f_d = 64$. 

Fig.~\ref{fig:snapshots_f2_and_f64_from_run1} indicates that the effect of \textit{trans} side friction and buckling on the waiting times $t_w$ depends on $f_d$. We study these effects via $n_d$, the number of beads causing friction, defined assuming overdamping in Sec.~\ref{sec:kappamapping} as $n_d(s)=[t_w(s) f_d]/(\xi m \sigma)$. To pin down effects coming from {\it trans} side, $n_d(s)$ for simulations where polymer beads are removed once they arrive at the \textit{trans} side (denoted as 'no {\it trans}') are compared with corresponding $n_d(s)$ for full polymers. We have used this bead removal previously for translocation of fully flexible polymers. Details of this procedure can be found in Refs.~\cite{Suhonen14}~and~\cite{Suhonen17}.

\begin{figure}[]
\begin{centering}
\includegraphics[width=1.00\linewidth]{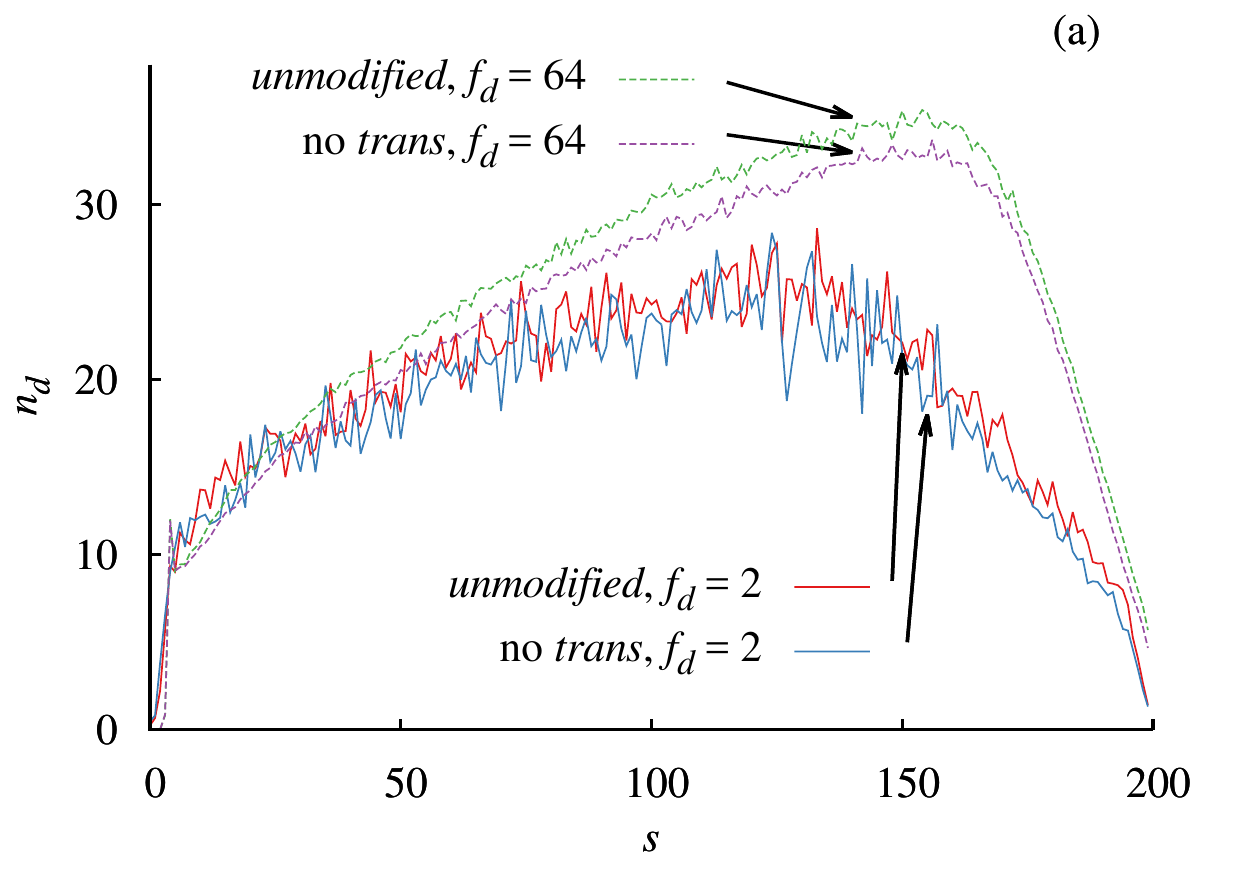}
\includegraphics[width=1.00\linewidth]{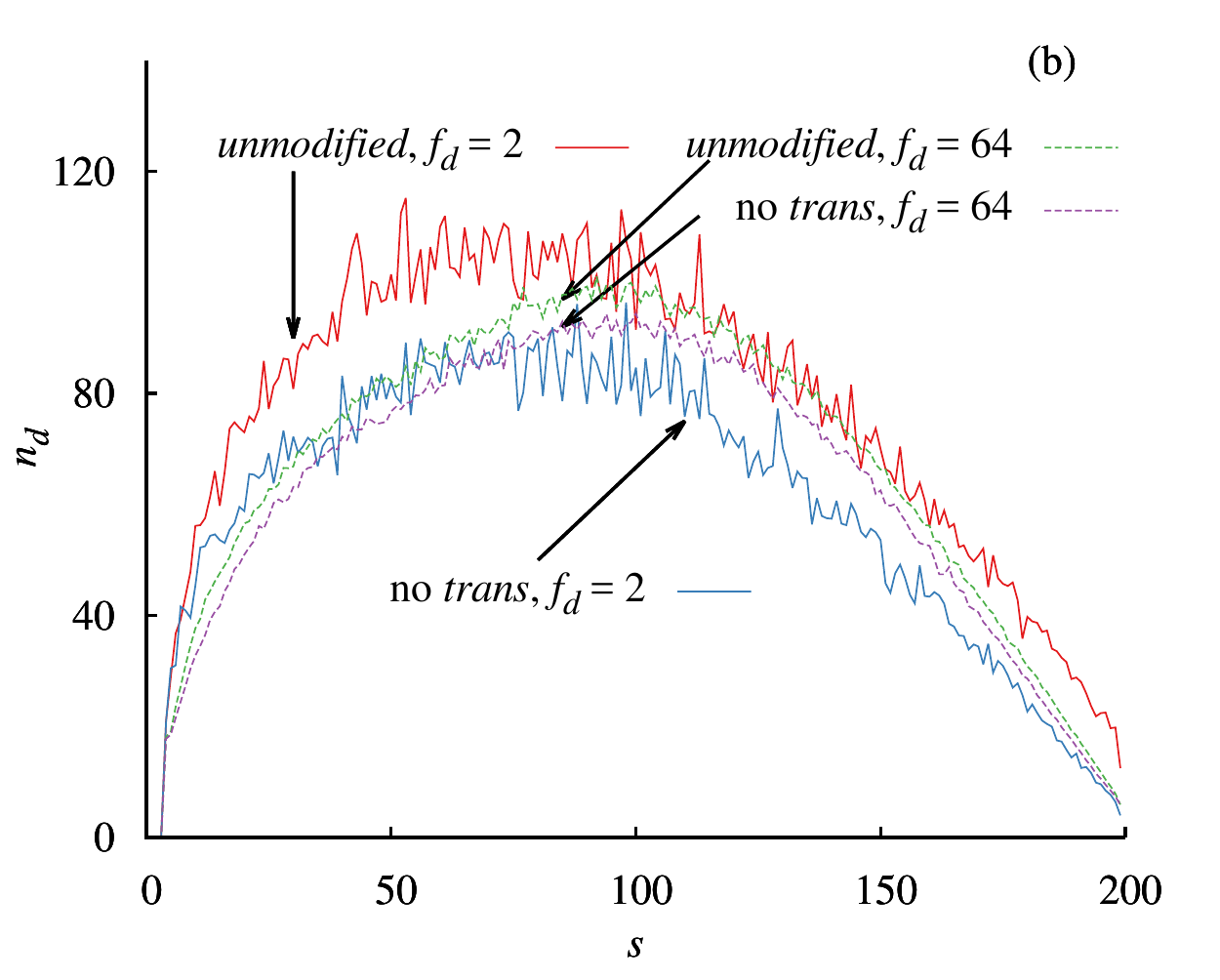}
\caption[]{(Color online) Effective $n_d$ calculated from $t_w(s)$ for \textit{unmodified} and no \textit{trans} translocation. $N = 200$. $f_d=2$ and $64$. (a) $\kappa = 0$ (b) $\kappa = 20$.}
\label{fig:normal_vs_notrans_f2etf64}
\end{centering}
\end{figure}

In Fig.~\ref{fig:normal_vs_notrans_f2etf64}~(a), $n_d(s)$ are shown for  fully flexible ($\kappa=0$) polymers of length $N=200$. $n_d(s)$ for \textit{unmodified} and no-{\it trans} polymers are seen to be essentially identical. Hence, there is no significant \textit{trans} side friction. Instead, the effect of diffusion can be seen clearly. The maximum $n_d(s)$ occurs at smaller $s$ for $f_d = 2$  than for $f_d = 64$. This is related to  the center of mass diffusion of the \textit{cis} side polymer segment~\cite{Suhonen17}. The center of mass diffusion releases tension that is seen to grow further in $s$ for $f_d = 64$, since the role of diffusion is small for such an extreme bias. Also, since the tensed segment on the {\it cis} side is shorter for $f_d=2$, $n_d(s)$ is smaller for $f_d=2$ than for $f_d = 64$.  

Having found clear center of mass diffusion also for polymers of $\kappa=20$ in Sec.~\ref{sec:diffusion}, we would expect $n_d$ curves for the semiflexible polymers to show similar dependence on $f_d$ as $n_d$ for fully flexible polymers. However, in Fig.~\ref{fig:normal_vs_notrans_f2etf64}~(b) $n_d(s)$ for \textit{unmodified} polymers of $\kappa=20$ is seen to be greater for $f_d=2$ than $f_d=64$. As seen in Fig.~\ref{fig:snapshots_f2_and_f64_from_run1}, the tensed segment on the {\it cis} side is longer for $f_d = 64$, so this difference of semiflexible and fully flexible polymers can only originate from the {\it trans} side.

Moreover, the difference in $n_d$ between \textit{unmodified} and no \textit{trans} polymers, which is significant for $f_d=2$, is quite small for $f_d=64$. This implies higher \textit{trans} side friction for smaller $f_d$. In Fig.~\ref{fig:snapshots_f2_and_f64_from_run1}  the \textit{trans} side segment is seen to buckle earlier for higher $f_d$. This buckling reduces friction, so the earlier buckling for higher $f_d$ would explain the observed effect. This characteristics is at odds with findings by Sarabadani et al., who found that {\it trans} side segment buckles more easily for low $f_d$~\cite{Sarabadani17}.

For comparison, we look at \textit{trans} side friction in the same way as was done in~\cite{Sarabadani17}. In other words, we compute from our simulations the geometrical quantity $\tilde{\eta}_{TS}$ that was used as a measure of \textit{trans} side friction in this paper. In order to calculate $\tilde{\eta}_{TS}$, a normalized angular cosine-correlation function is computed as $C(n)=\cos\delta_{s-1}\cos\delta_{s-2}\dots\cos\delta_{n}/\cos\delta_{s-1}$. Here angles $\delta_i$ from $i=s-1$ to $n$ refer to angles of polymer bond vectors with the normal of the membrane. Index $i=s-1$ refers to the latest bond that has arrived to the \textit{trans} side and $n \geq 1$. $\tilde{\eta}_{TS}$ is calculated as a sum $\tilde{\eta}_{TS}=\sum^{n^*}_{i=s-1}\cos\delta_i$, where $n^*$ is defined as the smallest index $i$ for which $C(i)<1/e$.

\begin{figure}[]
\includegraphics[width=0.49\linewidth]{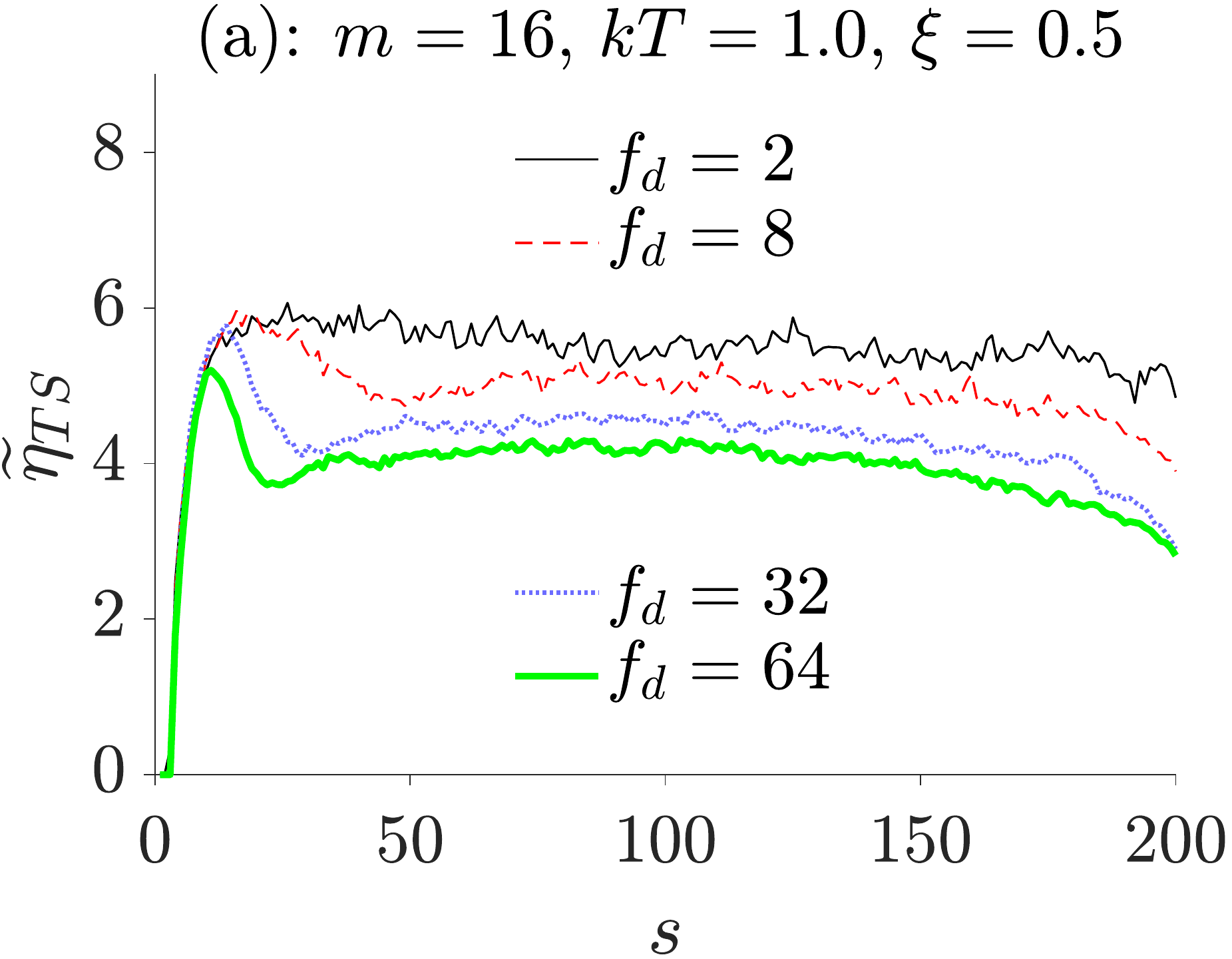}
\includegraphics[width=0.49\linewidth]{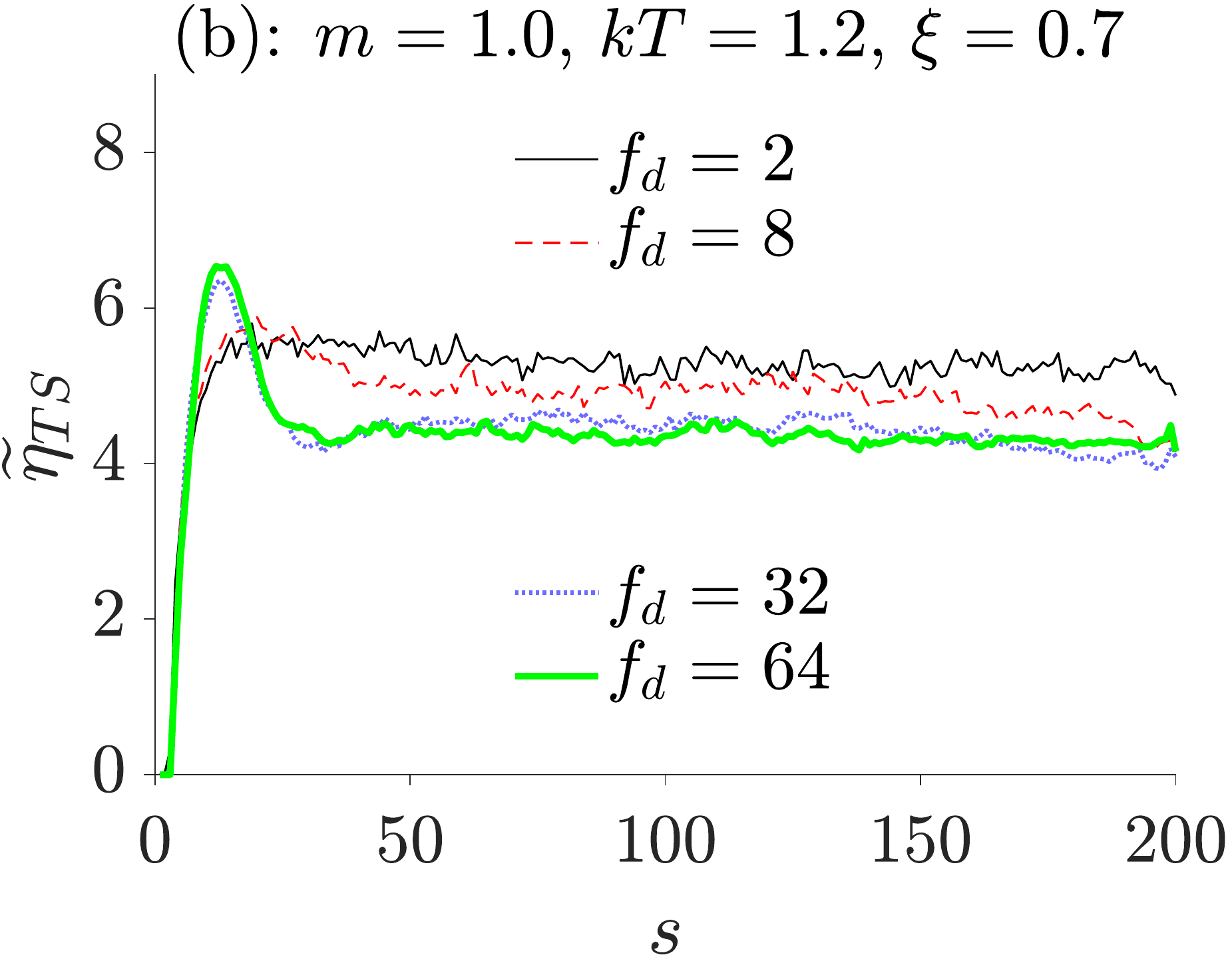}
\includegraphics[width=0.49\linewidth]{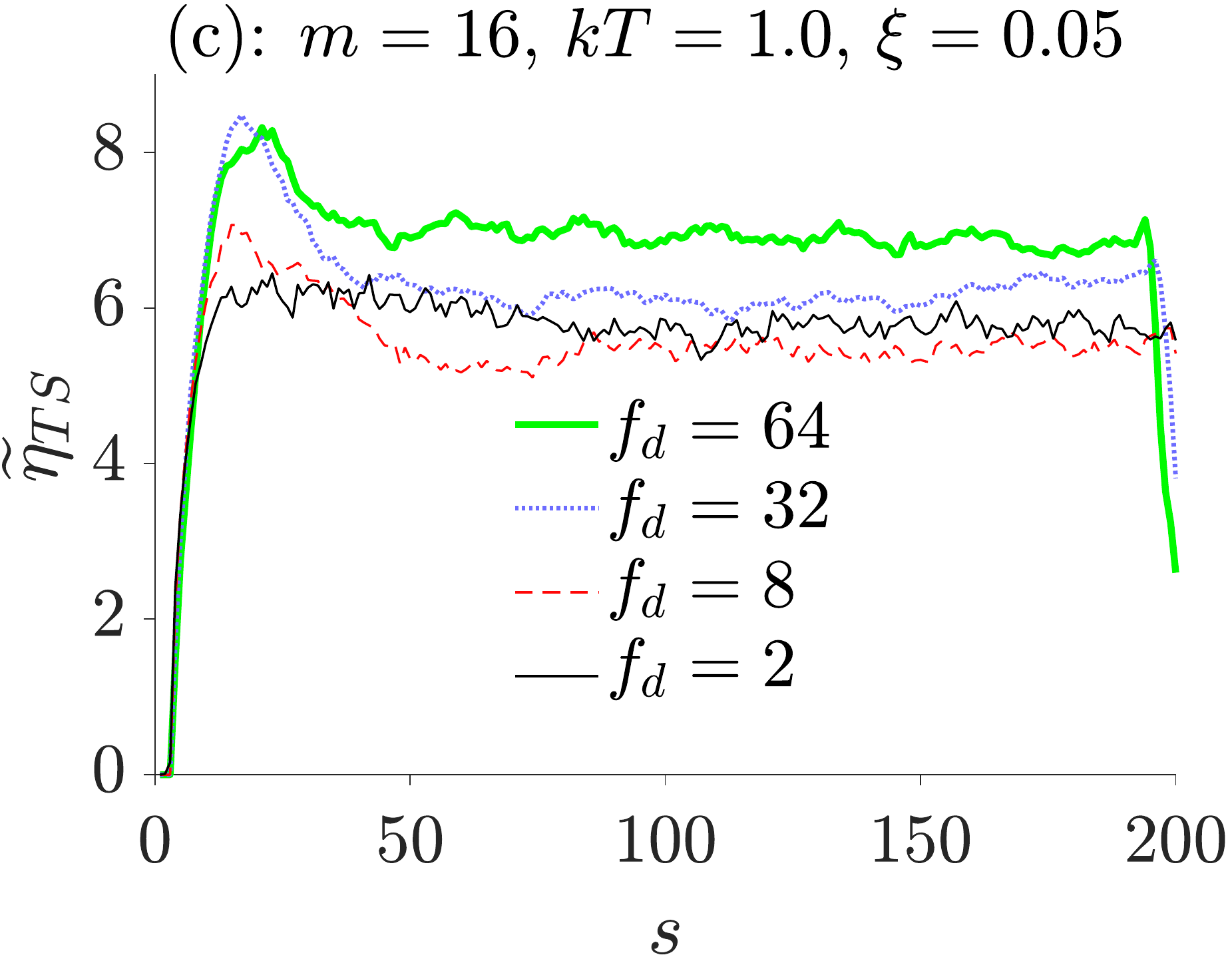}
\caption[]{(Color online) \textit{Trans} side friction contribution $\tilde{\eta}_{TS}$ calculated as in Ref.~\cite{Sarabadani17}. The figure corresponds to Fig.~3 of the supplemental material of Ref.~\cite{Sarabadani17}. For all (a), (b), and (c): $\kappa=20$ and $N=200$. (a) $\tilde{\eta}_{TS}$ is calculated from simulations with same parameters as used in this article (Parameters: $m=16$, $kT=1.0$, and $\xi=0.5$). (b) except for $\kappa$ and $N$, we use simulations with same parameters as in Ref.~\cite{Sarabadani17} (Parameters: $m=1$, $kT=1.2$, and $\xi=0.7$). This effectively reduces friction coefficient and increases diffusion coefficient. (c) $\xi$ is decreased to one tenth of what is used in (a) (Parameters: $m=16$, $kT=1.0$, and $\xi=0.05$).}
\label{fig:transfriction}
\end{figure}

In Fig.~\ref{fig:transfriction} we show $\tilde{\eta}_{TS}$ for $N=200$ and $\kappa = 20$ polymers with three different parameter sets. Fig.~\ref{fig:transfriction}~(a) shows $\tilde{\eta}_{TS}$ for our simulation parameters, where bead mass $m = 16$, temperature $kT=1.0$, and friction coefficient $\xi=0.5$. $\tilde{\eta}_{TS}$ is seen to be always lower for higher values of $f_d$. As the buckling reduces friction, this suggests easier buckling for higher $f_d$. For comparison, in Fig.~\ref{fig:transfriction}~(b) we use simulation parameters corresponding to those in~Ref~\cite{Sarabadani17}: $m=1$, $kT=1.2$, and $\xi=0.7$. At early stages of translocation, $\tilde{\eta}_{TS}$  is greater for higher $f_d$. After this, the same order as in Fig.~\ref{fig:transfriction}~(a) is resumed. Using these parameter values, we obtain qualitatively fairly similar characteristics for $\tilde{\eta}_{TS}$ as in~Fig.~3 of the supplemental material of Ref.~\cite{Sarabadani17}. However, in Ref~\cite{Sarabadani17}, the curves of different $f_d$ convergence to a same value as $s$ is increased. We do not observe this in either (a) or (b). Also, no sign of convergence is seen in Fig~\ref{fig:normal_vs_notrans_f2etf64}~(b), where the difference between \textit{unmodified} and no \textit{trans} models stays constant for both $f_d=2$ and $f_d=64$. Possibly the used smaller length of polymers explains the convergence observed in Ref~\cite{Sarabadani17}.

The proportion of dissipation to diffusion also affects the buckling. In translocation systems having weaker damping and higher diffusion, $\tilde{\eta}_{TS}$ can be higher for higher $f_d$. The dissipative term $-\xi m v_i$ in Eq.~(\ref{eq:langevin}) is greater in Fig.~\ref{fig:transfriction}~(a) than in Fig.~\ref{fig:transfriction}~(b). We think that this causes the easier buckling for low $f_d$ at small $s$ in Fig.~\ref{fig:transfriction}~(b). It is noteworthy that also the bead mass is different. For easier comparison, in Fig.~\ref{fig:transfriction}~(c), $\tilde{\eta}_{TS}$ is shown for simulations where $\xi$ is one tenth of $\xi$ in Fig.~\ref{fig:transfriction}~(a). The order of the curves is clearly reversed, and buckling becomes easier for lower $f_d$.

Most computer simulation models are more diffusive and less dissipative than those encountered in nature~\cite{deHaan15}. As we argued in~\cite{Suhonen17}, the value of the friction coefficient should ideally be higher than those generally used in computational models. Although generally the exact friction coefficient value does not seem to be of substantial importance as long as the dynamics remains in the dissipative regime, in buckling it seems to have relevance. Intuitively one would expect buckling to take place more easily for higher $f_d$. Based on the presented results we expect this to be the case in the relevant experimental translocation systems of higher viscosity and lower diffusion.

To slightly elaborate on the buckling, a beam on whose ends a force is exerted, buckles most likely in the middle. Probability of buckling also increases with the magnitude of the applied force. The situation is different for a semiflexible polymer chain being pushed from one end. Here, the drag force exerted on the chain is highest at the end where the force is exerted and decreases linearly along the chain. Hence, the drag force on the {\it trans} side is greatest close to the pore, where the buckling is most likely to happen. Also, in this case the probability for buckling is higher for larger force. The only factor that could change this is the thermal fluctuations of the heat bath. If the fluctuations are sufficiently strong compared to viscous drag force, then for weak force the fluctuations deviating the polymer backbone from the axis passing through the center of the axis might cause the buckling to be more probable for smaller force. However, this would require unrealistic low viscosity. In translocation the point at which the polymer segment on the {\it trans} side is pushed changes with $s$. The model of a rigid as yet unbuckled polymer being pushed at the end remains valid for all $s$.

\subsection{Scaling of translocation times $\tau$}\label{sec:scaling}

The average translocation time $\tau$ is known to obey the scaling $\tau \sim N^\beta f^\alpha$. In the very large $N$ and high $f_d$ regime $\beta=1+\nu$. For polymers of finite length and low $f_d$, $\beta<1+\nu$~\cite{Lehtola09,Rowghanian11,Ikonen12}. On the other hand for a completely straight chain $\beta=2$~\cite{Grosberg06}. It is thus reasonable to expect that including a bending potential would increase $\beta\ (\leq2)$.

In Table~\ref{tab:N_scaling} the scaling exponent $\beta$ obtained from our simulations is shown for $\kappa = 0$, $10$, and $20$, and $f_d = 2$, $32$, and $64$. The results show the expected increase of $\beta$ as a function of $\kappa$. It is also seen that $\beta$ increases as a function of $f_d$ for all $\kappa$. This characteristics, found previously for fully flexible chains~\cite{Lehtola09}, evidently holds also for semiflexible chains.

\begin{table}
\caption{Scaling exponents $\beta$ for the power law fits $\tau \sim N^\beta$. Calculated from $N=50$, $100$, $200$, and $400$.}
\begin{center}
\begin{tabular}{ |C{1.5cm}|C{1.5cm}|C{1.5cm}|C{1.5cm}| }
\hline
$f_d$  & $\kappa=0$ & $\kappa=10$& $\kappa=20$ \\
\hline
2 &  1.42&1.49  &1.52  \\
32 &  1.50&1.56  &1.60  \\
64 &  1.49&1.56  &1.61  \\
\hline
\end{tabular}
\end{center}
\label{tab:N_scaling}
\end{table}

The increase in $\beta$ as a function of $f_d$ is related to the center of mass diffusion in the low $f_d$ regime with fully flexible polymers~\cite{Suhonen17}. The center of mass diffusion towards the pore alleviates tension, speeding up translocation. As Fig.~\ref{fig:CMCis_wlc2} showed, this effect is present also for semiflexible polymers. For semiflexible chains there is also the $f_d$ dependent effect from \textit{trans} side friction. Unlike diffusion, additional \textit{trans} side friction slows down translocation for low $f_d$ thus reducing $\beta$. As seen in Table~\ref{tab:N_scaling} the combined effect of diffusion and \textit{trans} side friction results in increase in $\beta$ with $f_d$.

The \textit{trans} side friction also shows in the scaling exponent $\alpha$, see Table~\ref{tab:fd_scaling}. When for fully flexible polymers $\alpha>-1$, for semiflexible polymers $\alpha<-1$. This smaller $\alpha$ for semiflexible polymer translocation reflects the observed easier buckling of the {\it trans} side segment for stronger bias $f_d$ we found.

\begin{table}
\caption{Scaling exponents $\alpha$ for the power law fits $\tau \sim f_d^\alpha$. Calculated from simulations with $N=200$ and $f_d=2$, $32$, and $64$.}
\begin{center}
\begin{tabular}{ |C{1.5cm}|C{1.5cm}|C{1.5cm}| }
\hline
$\kappa=0$ & $\kappa=10$& $\kappa=20$ \\
\hline
-0.92&-1.02  &-1.04  \\
\hline
\end{tabular}
\end{center}
\label{tab:fd_scaling}
\end{table}

\begin{figure}[H]
\includegraphics[width=1.0\linewidth]{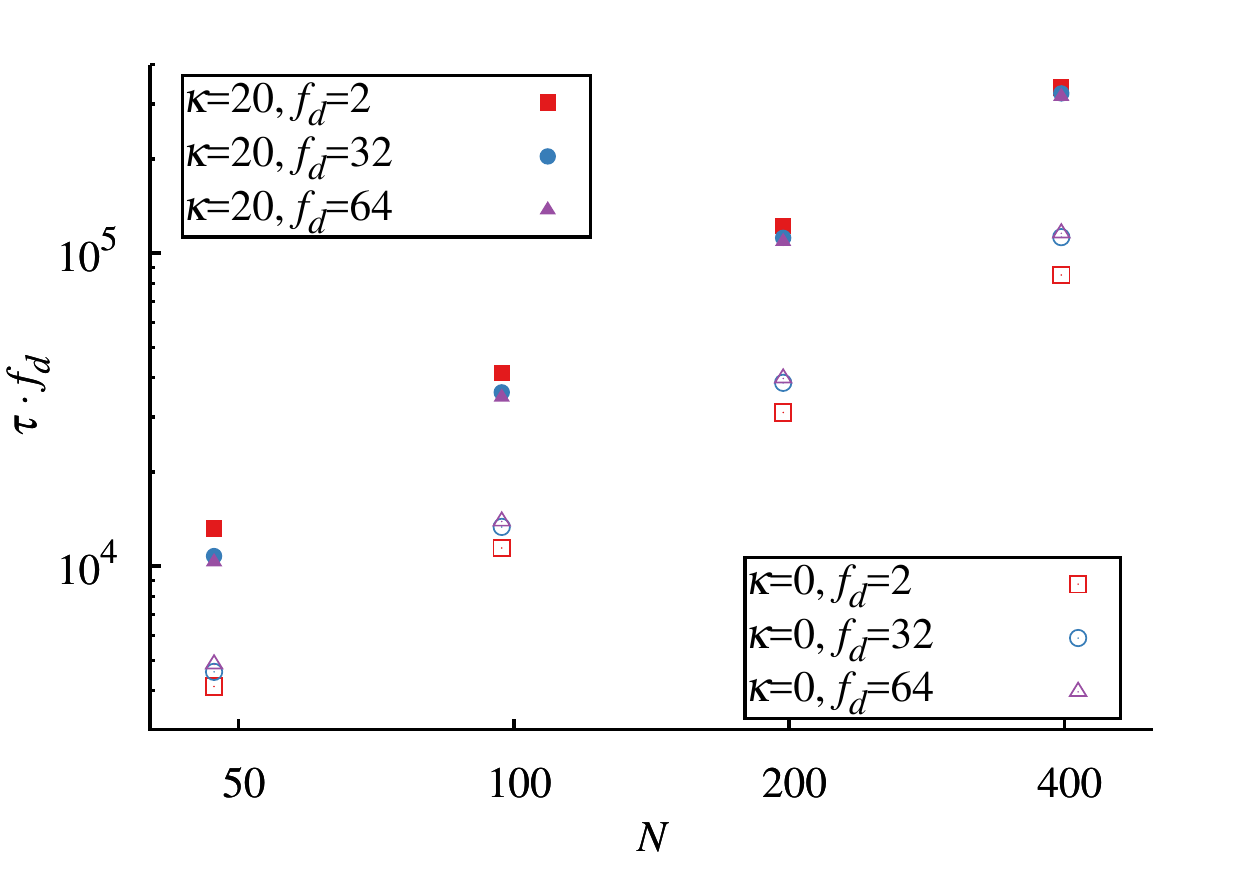}
\caption[]{(Color online) Scaled translocation times $\tau \cdot f_d$ as a function of $N$. Lower set of points: $\kappa = 0$, from bottom up $f_d = 2$, $ 32$, and $64$. The range of $N$ corresponds to $N_{\rm{Kuhn}} \in [28, 222]$. Upper set of points: $\kappa = 20$, from top down $f_d = 2$, $ 32$, and $64$. The range of $N$ corresponds to $N_{\rm{Kuhn}} \in [1.25, 10]$.}
\label{fig:tau_x_fd}
\end{figure}

To compare scaling of translocation time for different $f_d$, we plot scaled translocation times $\tau \cdot f_d$ vs $N$ for $\kappa=0$ and $\kappa=20$ in Fig.~\ref{fig:tau_x_fd}. The range of polymer lengths $N \in [50, 400]$ corresponds to $N_{{\rm Kuhn}} \in [28, 222]$ for fully flexible polymers and $N_{\rm{Kuhn}} \in [1.25, 10]$ for semiflexible polymers. For any $N$, $\tau \cdot f_d$  increases with $f_d$ for flexible polymers and decreases for semiflexible polymers. The increasing trend for flexible polymers is due to the more prominent contribution of the center-of-mass diffusion on the {\it cis} side for small $f_d$. The change of this trend for semiflexible polymers shows that {\it trans} side friction contributes to the dynamics.

For $\kappa=0$, the trend of $\tau \cdot f_d$ increasing with $f_d$ will be preserved for arbitrarily long polymers. When $\kappa=20$, the $\tau \cdot f_d$ curves for different $f_d$ will cross for sufficiently long polymers and will assume the same trend as the fully flexible polymers. This characteristics is reasonable. The \textit{trans} side friction is more significant for small $N_{\rm Kuhn}$. It is also  more pronounced for low $f_d$, when buckling does not occur as easily as for high $f_d$.

The {\it trans} side friction and buckling dynamics is seen to be relevant for polymers that are short in terms of their persistence length. It should be noted that in fact DNA of very finite lengths are used in experiments. For example, in experiments reported in Refs.~\cite{Storm05},~\cite{Fologea07},~and~\cite{Wanunu08} the longest dsDNA segments used were $97000$, $23000$, and $20000$ bp corresponding to $N_{\rm Kuhn} \approx 323$, $77$, and $67$ respectively. In the recent paper by Carson et al.~\cite{Carson14} the used dsDNA segments ranged from 35 to 20000 bp corresponding to $N_{\rm Kuhn} \approx 0.12$ to $N_{\rm Kuhn} \approx 67$. Sarabadani et al. fit their numerical model to this data~\cite{Sarabadani17}. The found scaling, $\tau \sim N^{1.37}$, is far from the scaling $\tau \sim N^{1.6}$ expected for asymptotically long polymers, as it should in the light of our findings. The dependence on $f_d$ is expected to be strong for the used polymer lengths. To claim an agreement between the model and experimentally found scaling comparison should be made for at least two $f_d$. In addition, the model Sarabadani et al. use has no hydrodynamic interactions, which is generally known to have a strong effect on the scaling $\tau \sim N^\beta$, see e.g.~\cite{Fyta08, Lehtola09, Moisio16}. Accordingly, the model and the experiments in~\cite{Sarabadani17} should not even in theory be in agreement.

\section{Conclusion}\label{sec:conclusion}

Here, we have studied driven translocation of semiflexible polymers through a nanoscale pore. We used Langevin dynamics simulations of a bead-spring polymer model including a bending potential characterized by rigidity parameter $\kappa$. The general characteristics of the process are aligned with expectations from physical considerations. Larger $\kappa$ resulted, due to increased friction, in slower translocation. Moreover, the scaling exponents $\beta$ in $\tau \sim N^\beta$ were found to increase as a function of $\kappa$, which is in keeping with $\beta \approx 1+\nu$ for fully flexible polymers and $\beta \approx 2$ for completely rigid polymers. In addition, $\beta$ was found to increase with $f_d$ for semiflexible polymers, similarly as has been found for fully flexible polymers~\cite{Lehtola09}.

To asses that driven translocation of semiflexible polymers shows unique characteristics not included in the respective process of fully flexible polymers, we made comparison by introducing the polymer length measured in the number of Kuhn segments $N_{\rm Kuhn}$. To make the discussion tangible, we somewhat simplistically transformed the waiting times to the fraction of polymers causing friction $\hat{n}_d/N$. We compared this quantity for fully flexible and semiflexible polymers, whose lengths were chosen such that the corresponding $N_{\rm Kuhn}$ agreed as closely as possible. The agreement was found to be fair but not perfect, especially for large $f_d$. Some of the differences can be attributed to differences in the excluded volume interactions and contribution from the effective pore friction, the magnitude of which is expected to change with changing $\kappa$. Even with these considerations, there clearly is a contribution from the {\it trans} side in driven semiflexible polymer translocation, which prevents treating semiflexible polymers as flexible polymers after a simple change in the relevant length scale for the used polymer lengths. 

In contrast to fully flexible polymers, where the \textit{trans} side effect is small~\cite{Suhonen14,Suhonen17}, we found the increased correlations of semiflexible polymers to cause a significant additional \textit{trans} side friction. We also found the intermittent buckling of semiflexible polymers to somewhat reduce this \textit{trans} side friction, in accord with findings of Sarabadani et al.~\cite{Sarabadani17}. However, in contrast to those findings we observed the buckling to take place more easily at high $f_d$. This causes the relative contribution of \textit{trans} side friction to be greater for smaller $f_d$ over the entire  process, again in contrast to Ref.~\cite{Sarabadani17}, where higher $f_d$ results in initially larger contribution, which then converges to the same friction for all $f_d$. This qualitative difference comes in one part from the greater dissipation present in our simulations and partly from polymers being longer in our Langevin simulations. The standardly used friction coefficients in computer simulations are generally small in comparison to real-world polymer systems~\cite{Suhonen17}. This is especially true for the crowded environment inside cells. In general, as long as the dynamics remains in the overdamped regime, the qualitative results do not deviate. However, we find that the buckling dynamics of semiflexible polymers is more sensitive to changes in the system's dissipation. The easier buckling of the polymer at large $f_d$ is intuitively expected and we showed that by increasing a system's dissipation, and so bringing it closer to realistic systems, this characteristics is obtained. At a more general level, in addition to {\it cis} side diffusion, also this {\it trans} side friction causes the scaling exponents to depend on the magnitude of the pore force.

We showed that, as in the case of fully flexible polymers~\cite{Suhonen17}, smaller $f_d$ results in significant center of mass diffusion toward the pore on the \textit{cis} side. This speeds up translocation of flexible polymers by alleviation of tension. For semiflexible polymers of low $N_{\rm Kuhn}$, this speed-up is smaller than the slowing down due to the \textit{trans} side friction. At some large enough $N_{\rm Kuhn}$, the roles are reversed and the situation becomes reminiscent to that of fully flexible polymers. However, DNA in translocation experiments are quite short in terms of their elastic rigidity. Hence, it is in fact dynamics of polymers of very finite lengths that is relevant. 

Our simulations here do not take into account hydrodynamic interactions. Hydrodynamics speeds up driven translocation of flexible polymers and decreases $\beta$~\cite{Fyta08,Lehtola09,Moisio16}. These effects are present also in the driven translocation of semiflexible polymers. Complete agreement of experimental measurements and computer models cannot be even theoretically expected if hydrodynamics is ignored and comparison is made only using one pore force magnitude.

\begin{acknowledgments}
Pauli Suhonen thanks Joonas Piili for useful discussions. The computational resources of CSC-IT Centre for Science, Finland, and Aalto Science-IT project are acknowledged. The work of Pauli Suhonen is supported by The Emil Aaltonen Foundation.
\end{acknowledgments}

\appendix

\bibliographystyle{ieeetr}
\bibliography{references.bib}

\end{document}